\documentclass[11pt,a4paper]{article}
\usepackage{url}
\usepackage{jcappub}
\usepackage{amsfonts}
\usepackage{subcaption}
\usepackage{graphicx}
\usepackage{appendix}

\usepackage[format=hang]{caption}
\usepackage{microtype}
\usepackage{epsfig}  
\usepackage{graphicx}   
\usepackage{slashed}       
\usepackage{tikz}
\usepackage{url}
\usepackage{color}
\usepackage{multirow}
\usepackage{comment}
\usepackage{amssymb}
\usepackage[capitalize]{cleveref}
\usepackage{bm}
\usepackage{soul}
\usepackage{orcidlink}
\usepackage[normalem]{ulem}
\usepackage{csquotes}
\usepackage{siunitx}

\usepackage{amsmath}
\usepackage{array}

\DeclareMathOperator{\cm}{cm}

\DeclareMathOperator{\MeV}{MeV}

\DeclareMathOperator{\s}{s}

\DeclareMathOperator{\km}{km}

\DeclareMathOperator{\g}{g}

\usepackage{scrextend}
\deffootnote[10pt]{10pt}{10pt}{\makebox[15pt][l]{\thefootnotemark.\hspace{10pt}}}

\newcommand{\gray}[1]{{\color{gray}{#1}}}

\definecolor{myblue}{rgb}{0.39, 0.58, 0.93}

\begin{document}

\title{Detecting Supernova Axions with IAXO}

\author[a]{P. Carenza~\orcidlink{0000-0002-8410-0345},}

\author[b,c]{J. A. García Pascual\footnotemark[1]~\orcidlink{0000-0002-7399-7353},}

\author[b,c,d]{M. Giannotti\footnotemark[1]~\orcidlink{0000-0001-9823-6262},}

\author[b,c]{I. G. Irastorza~\orcidlink{0000-0003-1163-1687},}

\author[b,c]{M. Kaltschmidt~\orcidlink{0000-0002-6470-5371},}

\author[e,f]{A. Lella~\orcidlink{0000-0002-3266-3154},}

\author[g]{A. Lindner~\orcidlink{0000-0001-6006-0820},}

\author[h]{G. Lucente~\orcidlink{0000-0003-1530-4851},}

\author[e,f]{A. Mirizzi~\orcidlink{0000-0002-5382-3786},}

\author[b,c]{M. J. Puyuelo~\orcidlink{0009-0005-7060-5341},}

\author[i]{T. Schiffer~\orcidlink{0009-0004-4746-8214}}

%To specify corresponding authors with asterisk
\renewcommand{\thefootnote}{\fnsymbol{footnote}}
\footnotetext[1]{Corresponding authors.}

\affiliation[a]{
The Oskar Klein Centre, Department of Physics, Stockholm University, Stockholm 106 91, Sweden}
\affiliation[b]{Centro de Astropart{\'i}culas y F{\'i}sica de Altas Energ{\'i}as (CAPA), Universidad de Zaragoza, Zaragoza, 50009, Spain}
\affiliation[c]{Departamento de Física Teórica, Universidad de Zaragoza, Zaragoza, 50009, Spain}
\affiliation[d]{Physical Sciences, Barry University, 11300 NE 2nd Ave., Miami Shores, FL 33161, USA}
\affiliation[e]{Dipartimento Interateneo di Fisica ``Michelangelo Merlin'', Via Amendola 173, 70126 Bari, Italy}
\affiliation[f]{Istituto Nazionale di Fisica Nucleare - Sezione di Bari, Via Orabona 4, 70126 Bari, Italy}
\affiliation[g]{Deutsches Elektronen-Synchrotron DESY, Notkestr. 85, 22607 Hamburg, Germany}
\affiliation[h]{SLAC National Accelerator Laboratory, 2575 Sand Hill Rd, Menlo Park, CA 94025}
\affiliation[i]{Physikalisches Institut der Universität Bonn, Nu\ss allee 12, 53115 Bonn, Germany}

\emailAdd{pierluca.carenza@fysik.su.se, juanangp@unizar.es, mgiannotti@unizar.es, Igor.Irastorza@cern.ch, mkaltschmidt@unizar.es, alessandro.lella@ba.infn.it, axel.lindner@desy.de, giuseppe.lucente@ba.infn.it, alessandro.mirizzi@ba.infn.it,  maria.jimenez@unizar.es, schiffer@physik.uni-bonn.de}

\abstract{
% \mg{I expanded the abstract a bit. Take a look.}
% We investigate the potential of IAXO and its intermediate version, BabyIAXO, to detect axions produced in core-collapse supernovae (SNe). 
% The study demonstrates that IAXO and BabyIAXO have realistic chances to detect supernova axions, potentially providing valuable insights into both axion physics and supernova dynamics. 
% We discuss the implementation of advanced gamma-ray detectors based on LiquidO technology to enhance detection efficiency. 
% The sensitivity of IAXO to supernova axions allows the exploration of axion parameter spaces inaccessible by solar observations and, in the case of a very close SN event and sufficiently large couplings, shed light on the axion production mechanisms in nuclear matter. 
We investigate the potential of IAXO and its intermediate version, BabyIAXO, to detect axions produced in core-collapse supernovae (SNe). Our study demonstrates that these experiments have realistic chances of identifying SN axions, offering crucial insights into both axion physics and SN dynamics. IAXO’s sensitivity to SN axions allows for the exploration of regions of the axion parameter space inaccessible through solar observations. 
In addition, in the event of a nearby SN, $d \sim \mathcal{O}(100)$ pc, and sufficiently large axion couplings, $g_{a\gamma}\gtrsim 10^{-11}\,{\rm GeV^{-1}}$, IAXO could have a chance to significantly advance our understanding of axion production in nuclear matter and provide valuable information about the physics of SNe, such as pion abundance, the equation of state, and other nuclear processes occurring in extreme environments. %Additionally, we explore the implementation of advanced gamma-ray detectors, such as those based on LiquidO technology, to enhance detection efficiency in the MeV energy range expected from SN axion emissions.
% \sout{This work highlights the scientific impact that a future Galactic SN could have for both particle physics and astrophysics, were axions to be detected.}
}

\maketitle
\date{\today}

\renewcommand{\thefootnote}{\arabic{footnote}}

\section{Introduction}
\label{sec:intro}

% \mg{Note that the list of authors is not the final one. The author's order should be decided later. I suggest JA and MG as corresponding authors. }
% \mg{Revised the intro. Please check.}

Axions~\cite{Weinberg:1977ma,Wilczek:1977pj} are theoretically well-motivated pseudo-scalar particles, predicted by several extensions of the Standard Model.
They are a robust prediction of the Peccei-Quinn (PQ) mechanism~\cite{Peccei:1977hh,Peccei:1977ur} of the strong CP problem, 
which remains one of the great challenges in particle physics.
Moreover, they are attractive dark matter (DM) candidates~\cite{Abbott:1982af,Dine:1982ah,Preskill:1982cy}. 
A quite extensive program is in place, worldwide, to detect these elusive particles. 
One of the most ambitious proposals is the International Axion Observatory (IAXO)~\cite{Irastorza:2011gs}, a new generation axion helioscope, 
which provides the sensitivity needed to probe a large section of the axion parameter space for a wide mass range. 
As a first step, the collaboration is building an intermediate-scale version, called BabyIAXO, 
which will function as a technological prototype for IAXO, as well as an independent physics experiment, with considerable discovery potential~\cite{IAXO:2020wwp}. 
%\GL{\textbf{Is there any reason why we never mention IAXO+ in Intro/conclusions/abstract but just in Sec.~\ref{sec:prospects}?}} 

The size of the IAXO and BabyIAXO magnets provides a large conversion volume which, combined with a novel tracking system  might permit the detection of an axion flux from a nearby supernova (SN), as originally proposed 
in Ref.~\cite{Raffelt:2011ft,Ge:2020zww}. 
Though the closest SN candidates are at $\mathcal{O}(100)$ pc from us, the very large temperature and density in the core of the exploding star could generate an axion flux large enough to be detected by a next generation helioscope. 
Specifically, for a sufficiently near candidate, the axion flux on Earth (in ${\rm cm^{-2}\,s^{-1}}$) might be (even significantly) larger than the expected solar axion flux.
The main difficulty, however, is that the SN axion flux drops rapidly after a few seconds, demanding readiness and a prompt intervention in case of such an event. 
Therefore, the \textit{helioscope as SN-scope} concept should rely on an efficient alert system for coming SN events.
The ability to anticipate a supernova event far enough in advance to prepare for observations is a topic of significant interest in current research.
A methodology for a pre-supernova alert has been already implemented in KamLAND~\cite{KamLAND:2015dbn},
and SNEWS 2.0 intends to develop and provide a pre-SN alert to inform of the coming CC event (see Sec. 4 of Ref.~\cite{SNEWS:2020tbu}). 
This capability practically relies on the detection of neutrinos emitted during the late evolutionary stages of a supergiant, particularly during the silicon-burning phase, which directly precedes the explosion (see Sec.~\ref{sec:IAXOSNScope} for more details).

% {\bf WARNING} An established alert system, the $\mathrm{SN}$ Early Warning System (SNEWS)~\cite{Antonioli:2004zb}, is already in place. 
% SNEWS provides an early warning for a Galactic core-collapse SN (CCSN) by measuring neutrinos from a soon-to-be SN and by tracking coincidences between neutrino experiments (see Sec.~\ref{sec:IAXOSNScope}). Its successor SNEWS 2.0~\cite{SNEWS:2020tbu} will reduce alert latency, providing even faster notifications of potential supernova events. 
% {\bf AM: WARNING: I AM NOT SURE SNEWS IT WOULD BE USEFUL. SNEWS IS BASED ON SN NEUTRINOS WHICH WILL ARRIVE TOGETHER WITH AXIONS. INSTEAD, ONE SHOULD DETECT PRE SN NEUTRINOS FROM SI BURNING. I AM NOT SURE SNEWS HAS A ROLE SINCE MANY NU DETECTORS IN SNEWS WILL NOT OBSERVE PRE SN. SURELY NONE OF THE CURRENT ONES. SO THIS IS MORE A PERSPECTIVE FOR THE FUTURE}\mg{Yes. Perhaps, better to remove the discussion on SNEWS, which we do not use anyway. In the end we consider only the results from Cecila Lunardini et al. to give quantitative ideas about the pre-SN signal}

% The SuperNova Early Warning System and the 
The fast response time implemented in the IAXO and BabyIAXO helioscopes
% , based on that of CTA (Cherenkov Telescope Array) telescopes, 
facilitates the implementation of IAXO as a SN-scope. 
In contrast to helioscope searches, where the expected axion signature is X-rays in the keV regime, the different SN production mechanisms lead to axion energies in the $10-250~\mathrm{MeV}$ energy range. Therefore, the detection of SN axions relies on the development of a highly efficient gamma-ray detector in addition to the X-ray detector. In particular, the gamma-ray detector should provide topological information to distinguish the signature of high energy axions from background events.

Given the current technology capabilities, the detection of axions from a nearby SN event is a realistic possibility and the construction of a dedicated MeV detector for the next generation of axion helioscopes is highly motivated.
First, a sensitivity to SN axions would allow us to explore regions of the axion parameter space not accessible through solar observations. 
Second, such a detection could shed light on important aspects of the SN physics, not accessible through other experimental tests.
SNe are very rare events
with current estimates predicting a rate of $1.63 \pm 0.46$ per century~\cite{Rozwadowska:2020nab}. 
% We should be sure to be as prepared as possible for such a prospect.
It is, therefore, imperative to ensure the highest level of preparedness for such an eventuality.

In this work, we focus on the theoretical and phenomenological aspects of the helioscope potential to detect axions from a nearby SN event. 
The discussion of the most appropriate MeV detector configuration for IAXO is presently ongoing and will be presented in a separate work.

This article is organized as follows. In Section \ref{sec:axion_prod} we summarize the different SN axion production mechanisms. 
Section \ref{sec:IAXOSNScope} describes the different technical aspects for the implementation of IAXO and BabyIAXO for the detection of SN axions. In Section \ref{sec:prospects} the sensitivity prospects in different scenarios are presented. Finally, we conclude our discussion of IAXO as a SN-scope in Section \ref{sec:conclusion}. There follow Appendices with more technical details of our analysis.

\section{Supernova axion production}
\label{sec:axion_prod}
% \red{Maurizio, Alessandro Lella, Mirizzi, Pierluca, Giuseppe}\\

Core-collapse SNe are among the most powerful astrophysical factories for the production of feebly-interacting particles. Indeed, the series of catastrophic events occurring during the onset of the gravitational collapse induces extreme conditions of temperature and density in the inner regions of the core (see Refs.~\cite{Janka:2006fh,Mirizzi:2015eza,Carenza:2024ehj} for some reviews). In particular, at the beginning of the cooling phase the core of an exploding SN is expected to be characterized by temperatures $T\sim30-40\,\MeV$ and densities around the nuclear saturation $\rho\sim10^{14}\,\g\,\cm^{-3}$. Under these conditions, the production of feebly-interacting particles, like axions and axion-like particles (ALPs), could be significantly enhanced. Therefore, a future galactic SN event may represent a once-in-a-lifetime opportunity to probe axion physics.

\begin{figure} [t!]
\centering
\includegraphics[width=0.42\textwidth]{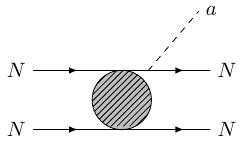}
\hfill
\includegraphics[width=0.42\textwidth]{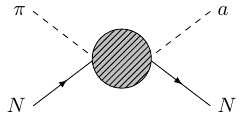}
\caption{Schematic representation of the Feynman diagrams for the $NN$ bremsstrahlung (\textit{left panel}) and the pionic Compton-like process (\textit{right panel}).}
\label{fig:NuclearProcesses}
\end{figure}

If axions are coupled to nuclear matter, the main production channels in the hot and dense SN core depend on their coupling to nucleons $g_{aN}$, with $SN axionN=n,\,p$ for neutrons and protons respectively~\cite{Lella:2022uwi,Lella:2023bfb,Lella:2024dmx}. The first considered channel for axion emission from the SN nuclear medium is $NN$ bremsstrahlung $N+N \rightarrow N+N+a$~\cite{Carena:1988kr,Brinkmann:1988vi,Turner:1991ax,Raffelt:1996wa,Keil:1996ju}. As depicted in the left panel of Fig.~\ref{fig:NuclearProcesses}, in this process an axion is produced as a consequence of strong interactions between two nucleons in the medium. The state-of-the-art calculation for the bremsstrahlung emission rate has been performed in Ref.~\cite{Carenza:2019pxu} and accounts for corrections beyond the usual one-pion exchange (OPE) approximation for the nuclear interaction potential. In particular, a non-vanishing mass for the exchanged pion~\cite{Stoica:2009zh}, the contribution from the two-pion exchange~\cite{Ericson:1988wr}, effective in-medium nucleon masses and multiple nucleon scattering effects~\cite{Raffelt:1991pw,Janka:1995ir} have to be taken into account to obtain a reliable estimation of the bremsstrahlung emissivity. Finally, Ref.~\cite{Springmann:2024mjp} has discussed the impact of finite-density effects over ALP-nucleon couplings in dense environments.
Furthermore, the right panel of Fig.~\ref{fig:NuclearProcesses} displays a schematic representation of pionic Compton-like processes $\pi+N \rightarrow a+N$, in which a pion is converted into an axion after the scattering onto a non-relativistic nucleon. Since in the SN nuclear medium the fraction of thermal pions $Y_\pi$ was expected to be extremely small $Y_\pi\sim\mathcal{O}(10^{-4})$~\cite{Caputo:2024oqc}, the contribution from pion conversions has been overlooked for a long time. Nevertheless, the authors of Ref.~\cite{Fore:2019wib} pointed out that strong interactions can magnify the abundance of negatively-charged pions. Thus, the pion conversion emission rate has been re-estimated in Ref.~\cite{Carenza:2020cis}, realizing that, under the assumption of Ref.~\cite{Fore:2019wib}, it could be comparable and even dominant over the bremsstrahlung contribution. Moreover, it has been recently shown that the contribution from the contact interaction term~\cite{Choi:2021ign} and the $\Delta(1232)$ resonance~\cite{Ho:2022oaw} can significantly enhance axion emissivity via pion conversion. We address the reader to Refs.~\cite{Lella:2022uwi,Carenza:2023lci} for the complete expressions of both $NN$ bremsstrahlung and pion conversion emission rates. 

%%%%%%%%%%%%%%%%%%%%%%%%%%%%%%%%%%%%%%%%%%%%
%%%%%%%%%%%%%%%%%%%%%%%%%%%%%%%%%%%%%%%%%%%%

%%%%%%%%%%%%%%%%%%%%%%%%%%%%%%%%%%%%%%%%%%%%
\begin{figure} [t!]
\centering
\includegraphics[width=0.7\textwidth]{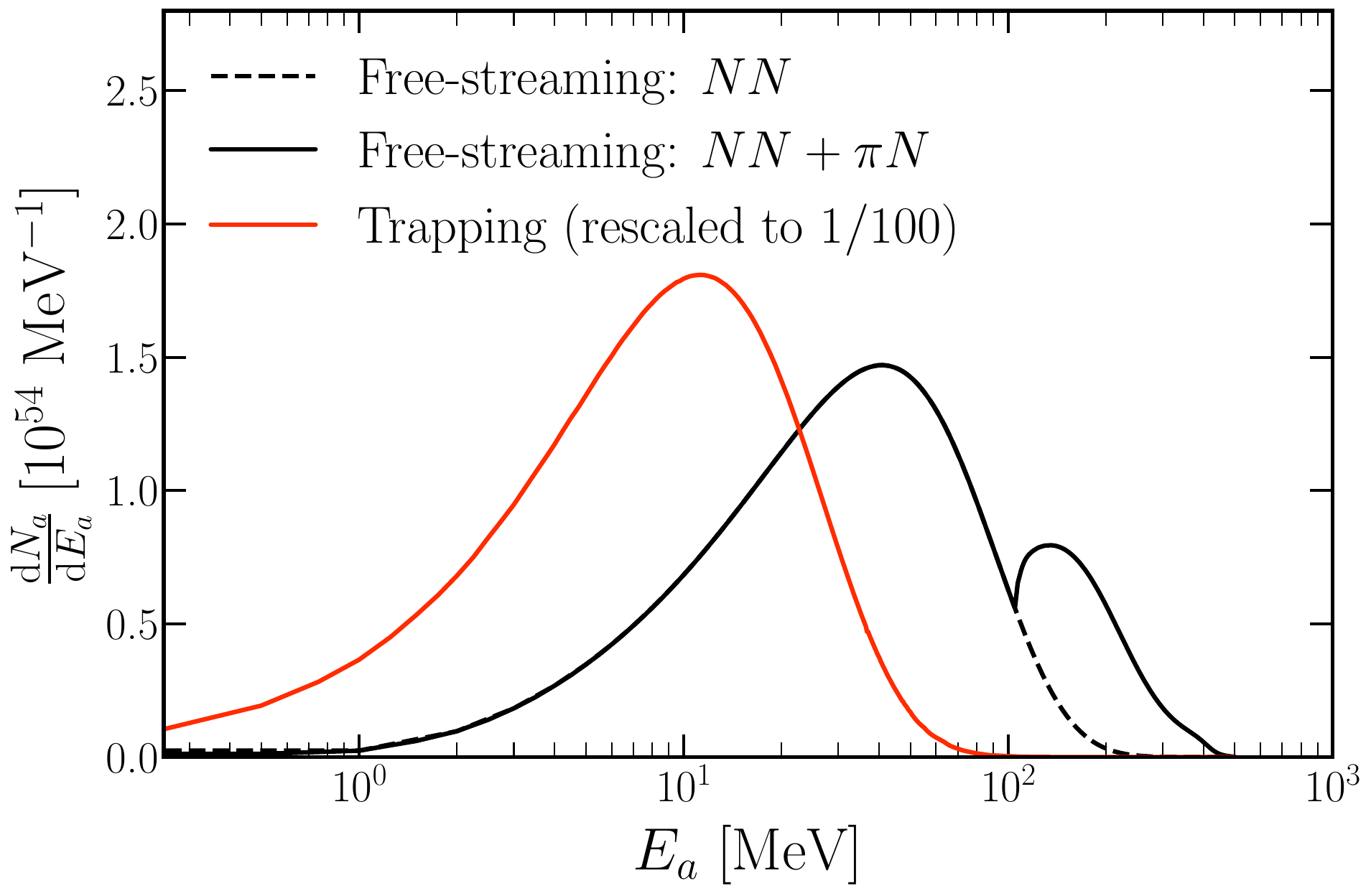}
    \caption{Time-integrated ALP emission spectrum from our benchmark SN model \cite{SNarchive}. The continuous black line depicts the spectrum line-shape including both contributions from $NN$ bremsstrahlung ($NN$) and pion conversions ($\pi N$), while the dashed black line refers to the bremsstrahlung-only contribution in the \textit{free-streaming regime}. The corresponding spectrum for the \textit{trapping regime} is depicted as a continuous red line. Note that we rescaled the trapping curve by a factor of 1/100 for the visualization. Here the ALP-neutron coupling is fixed at $g_{an}=0$, while the ALP-proton coupling is set at $g_{ap}=5\times10^{-10}$ free streaming scenario and at $g_{ap}=3\times10^{-6}$ in the trapping case.
    % {\bf why not log scale on x- axis?}
    }
\label{fig:AxSpectra}
\end{figure}
%%%%%%%%%%%%%%%%%%%%%%%%%%%%%%%%%%%%%%%%%%%%

If axions are weakly coupled to nuclear matter, ${g_{aN}\lesssim10^{-8}}$, they are in the \textit{free-streaming regime}, in which they can leave the star without being reabsorbed by the nuclear medium in the SN core. 
Therefore, they can act as an additional energy-loss channel during the SN cooling phase, contributing to the shortening of the neutrino burst duration~\cite{Raffelt:1987yt}. 
In particular, consistency with observations of the neutrino burst from SN 1987A\footnote{Current state-of-the-art SN simulations, including convection and updated neutrino-nucleon opacities, are in tension with late-time SN 1987A events, since they predict a cooling time shorter than the observed one~\cite{Fiorillo:2023frv}.} requires ${g_{ap}\lesssim5\times10^{-10}}$~\cite{Lella:2023bfb}. 
On the other hand, for sufficiently large couplings $g_{aN}\gtrsim10^{-8}$, the SN core becomes ``optically thick'' for the produced axions, so that they can be also reabsorbed
inside the core by means of the inverse processes $N+N+a\rightarrow N+N$ and $a+N\rightarrow\pi+N$~\cite{Lella:2023bfb}. 
In this regime, dubbed \emph{trapping regime}, axions are not able to escape the inner regions of the core and their emission can be approximately described as a blackbody radiation from a last-scattering surface~\cite{Caputo:2022rca}, in close analogy with the case of neutrinos. We highlight that a complete treatment of the trapping regime would require the self-consistent addition of axions in SN simulations, since they could significantly impact the SN evolution. Relying just on the SN cooling argument, the range $5\times10^{-10} \lesssim g_{ap}\lesssim3\times10^{-6}$ is excluded. However, larger values of the coupling can be constrained by the non observation of axion-induced events in Kamiokande-II during SN 1987A~\cite{Engel:1990zd,Lella:2023bfb}.
Nevertheless, despite these constraints, even the trapped region remains a subject of interest for axion experiments~\cite{Giannotti:2024xhx,Arias-Aragon:2024gdz,Carenza:2023wsm,Alonso-Gonzalez:2024ems,Lella:2023bfb,Carenza:2024ehj}.
% \sout{the limited statistics and sparse data from the SN 1987A neutrino burst highlight the need for further exploration of these regions of the axion parameter space.}
%{\bf The last sentence is somehow a weak motivation. In case of strongly coupled axions we would have found many events, not a sparse sample. Since we have few events probably we are pretty sure that there are not trapped axions.}
%\mg{I feel similarly. Should we remove the last stentece? }

In the following, we adopt as benchmark SN model the 1D spherical-symmetric {\tt GARCHING} group's SN model SFHo-s18.8 provided in Ref.~\cite{SNarchive}, used also in previous studies about SN axions (see, e.g. Refs.~\cite{Bollig:2020xdr,Caputo:2021rux,Caputo:2022mah,Lella:2022uwi,Lella:2023bfb,Lella:2024hfk,Lella:2024dmx}). The simulation, based on the neutrino-hydrodynamics code 
{\tt PROMETHEUS-VERTEX}~\cite{Rampp:2002bq}, is launched from a stellar progenitor with mass $18.8~M_\odot$~\cite{Sukhbold:2017cnt} and leads to a neutron star (NS) with baryonic mass $1.35~M_\odot$, a typical value expected for a SN explosion event~\cite{Janka:2006fh}. 
More details on the employed SN model are provided in  Appendix~\ref{app:SNmodels}. The presence of pions in the SN core has been taken into account by following the procedure illustrated in Ref.~\cite{Fischer:2021jfm}, including the pion-nucleon interaction as described in Ref.~\cite{Fore:2019wib}.
We highlight that the SN model employed in this work is characterized by relatively low SN core temperatures compared to the other profiles provided in Ref.~\cite{SNarchive}~(see, e.g., Ref.~\cite{Caputo:2021rux,Manzari:2024jns}). Therefore, it is expected to provide the most conservative results for the SN axion fluxes. A discussion of typical uncertainties related to SN limits on axion couplings is reported in Ref.~\cite{Lella:2023bfb}.
Figure~\ref{fig:AxSpectra} shows the axion emission spectrum integrated over $10\,\s$ in both the free-streaming (black lines) and trapping regimes (red line) for our benchmark SN model. In the free-streaming regime, the spectrum is characterized by a bimodal shape, since the emission peak associated with the $NN$ bremsstrahlung is located at axion energies $E_a\sim50\,\MeV$, while the pion conversion contribution peaks at $E_a\sim200\,\MeV$~\cite{Lella:2022uwi,Lella:2023bfb}. 
This is related to the different nature of the two processes. In particular, the bremsstrahlung behaves as a quasi-thermal process, so that the average energy of the emitted ALP can be estimated as $E_a\simeq\frac{3}{2}T$ ~\cite{Lella:2024hfk}. 
On the other hand, the pion conversion is highly non-thermal and the average ALP energy is given by $E_a\simeq\,m_\pi+\frac{3}{2}T$, where $m_\pi\simeq135\,\MeV$ is the pion mass.

By summing the contributions from both nuclear processes,
the axion emission spectrum is expected to have the form~\cite{Lella:2022uwi,Lella:2023bfb}
%%%%%%%%%%%%%%%%%%%%%%%%%%%%%%%%%%%%%%%%%%%%
\begin{align}
\label{eq:emittd_axion_spectrum-FreeStreaming}
    \frac{dN_a}{dE_a}= A\,g_{aN}^2\left[ F_{NN}(E_a)+\delta\, F_{\pi N}(E_a)\right]\,.
\end{align}
%%%%%%%%%%%%%%%%%%%%%%%%%%%%%%%%%%%%%%%%%%%%
Here, $A$ is a normalization factor, with the dependence on the axion coupling constant factorized for convenience.
The functions $F_{NN}$ and $F_{\pi N}$ encode the spectral shape of bremsstrahlung and pionic processes, respectively.
These functions, as well as $A$, depend on the specific SN model, in particular on the equation of state employed and on the specific plasma conditions in the SN nuclear medium (simple fitting expressions for axion spectra, considering a range of supernova simulations, can be found in Ref.~\cite{Lella:2024hfk}).
Finally, $\delta$ is a free parameter that quantifies the relative contributions of $F_{NN}$ and $F_{\pi N}$ with respect to the benchmark model (corresponding to $\delta=1$ and shown in Fig.~\ref{fig:AxSpectra}). 
Therefore, $\delta$ serves as an indicator of the pion abundance in the proto-neutron star (PNS), normalized so that $\delta=1$ corresponds to a pion-to-nucleon fraction $Y_\pi \sim \mathcal{O}(1)\%$ in the inner core regions (see Appendix~\ref{app:SNmodels} for further details).

In the trapping regime, axions are emitted from outer regions of the SN core at ${R\sim15-20\,\km}$ where the temperature is in the range of few MeV. Under these conditions, the pion conversion peak is completely suppressed, while the average energy of ALPs produced by $NN$ bremsstrahlung processes is shifted towards lower values $E_a\sim10\,\MeV$~\cite{Lella:2023bfb}.

\section{Direct detection of SN axions in IAXO}
\label{sec:IAXOSNScope}
% \red{Maurizio, JuanAn, Mathieu}

% The novel tracking system for new generation axion helioscopes is based on that of Cherenkov Telescope Array (CTA) telescopes \cite{CTA:2013introducing} and offers the possibility to point at a specific target with a relatively fast response time ($< 90\,\s$).
The novel tracking system for new generation axion helioscopes offers the possibility to point at a specific target within a few minutes~\cite{IAXO:2020wwp}.
We plan to take advantage of this system to detect SN axions with IAXO, by orienting it toward the SN. 
In this section we present a detailed and realistic analysis for the detection possibilities with IAXO.

\begin{figure}[t]
    \centering
    \includegraphics[width=1\textwidth]{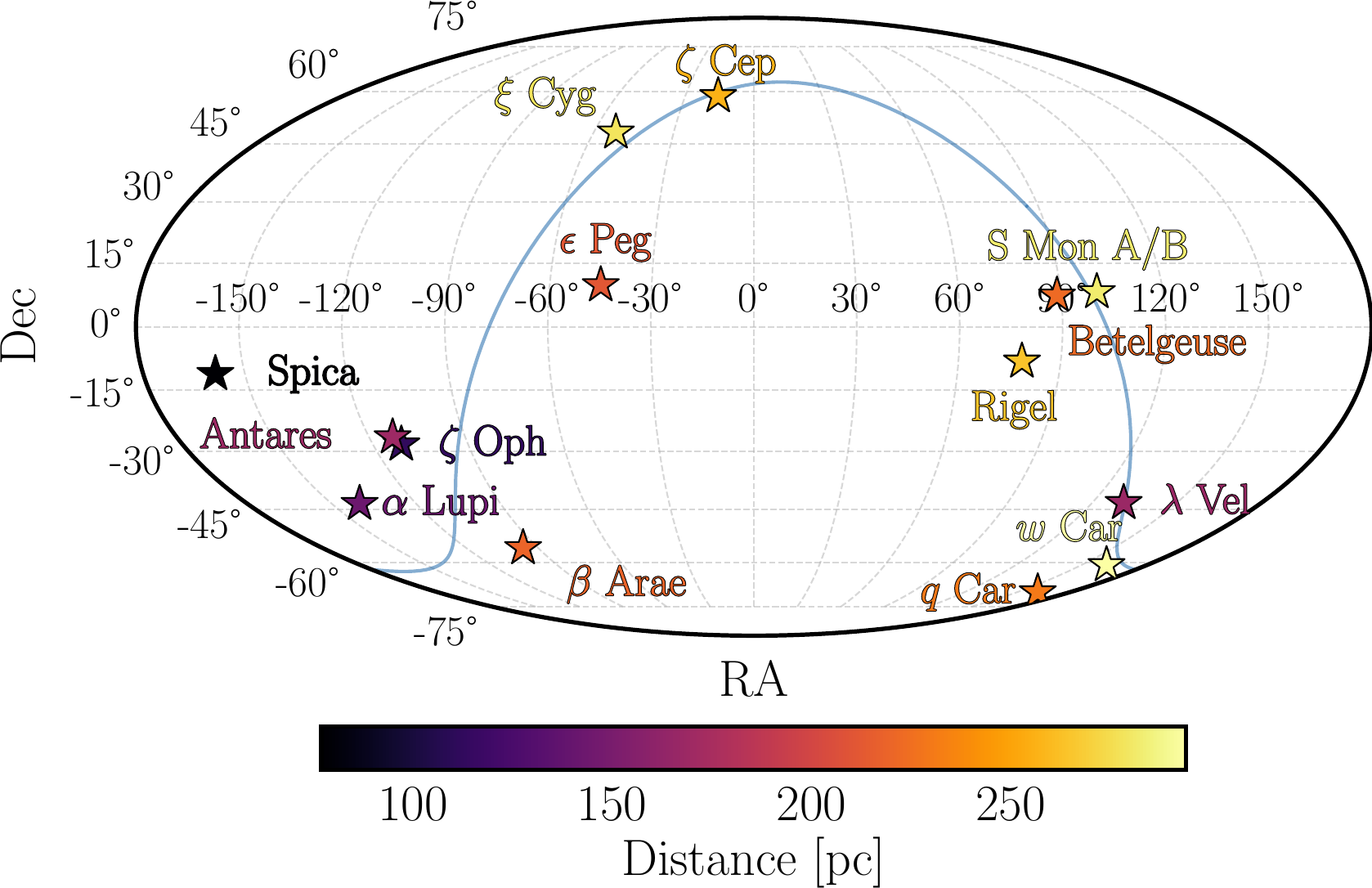}
    \caption{Mollweide projection
    % (in the ICRS frame) \GL{Define ICRS} 
    of the SN candidates presented in Tab.~\ref{tab:SN-Candidate}.  The Galactic plane, defined by latitude 0 in the Galactic coordinate system, is depicted as a continuous blue line across the sky. 
    The color scheme encodes the distance of the stars from Earth in parsec.}
    \label{fig:mollweide}
\end{figure}

The list of nearby SN candidates is presented in Table ~\ref{tab:SN-Candidate}. 
A corresponding Mollweide projection of the position of the listed candidates in the galactic plane is subsequently presented in Figure \ref{fig:mollweide}. 
Examples of the error expected in the directional determination for three selected candidates are shown in Tab.~\ref{tab:B-IAXO_pos}. 
The data is available in Ref.~\cite{Mukhopadhyay:2020ubs}, though we have integrated it with the more recent red supergiant catalog presented in Ref.~\cite{Healy:2023ovi}\footnote{Effectively, the integragion with the Healy et al. catalog~\cite{Healy:2023ovi} resulted in the addition of the star $\beta$ Arae ($N$=7) to the list of nearby supergiants in Ref.~\cite{Mukhopadhyay:2020ubs}.} and crosschecked with the \texttt{simbad} database \cite{Wenger:2000sw}.
We limit our list to $d<300$ pc as the axion flux from more distant stars would be too small to be detected with IAXO.

\begin{table}[t!]
\renewcommand{\arraystretch}{1.5}
\centering

\begin{tabular}{cccclc}
\hline
\hline
$N$  & Catalog Name & Common Name                 & Distance [kpc]      & RA  [J2000]        & Dec [J2000]           \\ \hline
1  & HD 116658    & Spica/$\alpha$ Virginis     & 0.077 $\pm$ 0.004   & 13:25:11.58 & -11:09:40.8           \\ %\hline
2  & HD 149757 & $\zeta$ Ophiuchi  & 0.112 $\pm$ 0.002 & 16:37:09.53 & -10:34:01.5  \\ %\hline
3  & HD 129056 & $\alpha$ Lupi     & 0.143 $\pm$ 0.003 & 14:41:55.79 & -47:23:17.52 \\ %\hline
4  & HD 78647  & $\lambda$ Velorum & 0.167 $\pm$ 0.003 & 09:02:06.85 & -43:31:17.72 \\ %\hline
5  & HD 148478    & Antares/$\alpha$ Scorpii    & 0.169 $\pm$ 0.003   & 16:29:24.46 & -26:25:55.2           \\ %\hline
6  & HD 206778 & $\epsilon$ Pegasi & 0.211 $\pm$ 0.006 & 21:44:11.16 & +09:52:30.0  \\ %\hline
7  & HD 157244    & $\beta$ Arae                & $0.219 \pm 0.010$ &   17:25:17.99          & -55:31:47.57 \\ %\hline
8  & HD 39801     & Betelgeuse/$\alpha$ Orionis & 0.222 $\pm$ 0.040   & 05:55:10.31 & +07:24:25.4           \\ %\hline
9  & HD 89388  & q Car/V337 Car    & 0.230 $\pm$ 0.022 & 10:21:20.75 & -61:48:29.2  \\ %\hline
10 & HD 210745 & $\zeta$ Cephei    & 0.256 $\pm$ 0.006 & 22:10:51.23 & +58:12:04.4  \\ %\hline
11 & HD 34085     & Rigel/$\beta$ Orionis       & 0.264 $\pm$ 0.024   & 05:14:32.27 & -08:12:05.9           \\ %\hline
12 & HD 209005 & $\zeta$ Cygni     & 0.278 $\pm$ 0.029 & 21:19:51.82 & +30:13:49.2  \\ %\hline
13 & HD 47839  & S Monocerotis A   & 0.282 $\pm$ 0.040 & 06:40:58.66 & +09:53:44.7  \\ %\hline
14 & HD 47839  & S Monocerotis B   & 0.282 $\pm$ 0.040 & 06:40:58.66 & +09:53:44.7  \\ %\hline
15 & HD 93070  & q Car/V520 Car    & 0.294 $\pm$ 0.023 & 10:43:32.49 & -60:06:29.0  \\ \hline\hline
\end{tabular}
\caption{
List of all SN candidates at distance $d\leq 300$ pc. 
The catalog name refers to the Henry Draper (HD) catalog number. 
Data from Refs.~\cite{Mukhopadhyay:2020ubs,Healy:2023ovi} and the \texttt{simbad} database \cite{Wenger:2000sw}. 
}
\label{tab:SN-Candidate}
\end{table}

It is evident that the detection of SN axions using IAXO and BabyIAXO relies heavily on our ability to predict core-collapse SN events with sufficient advance notice and angular precision to plan observations effectively.
For this purpose, one has to rely on an early-alert system.
Effectively, this is possible through the observation of pre-SN neutrinos, as
the neutrino production rises  rapidly during the last stages of nuclear burning~\cite{Patton:2017neq}. 
Although the flux remains substantially lower—and at lower energies—than what is expected during the explosion itself, these neutrino fluxes are predicted to be observable in dedicated detectors minutes to days before the rise of the electromagnetic emission, providing the desired early alert (see, e.g., Ref.~\cite{Patton:2017neq}). 
Dedicated analyses indicate, for example, that the combined analysis of KamLAND and Super-Kamiokande (SK) can detect the pre-SN neutrino flux from stars up to $\sim$ 500 pc, providing an early warning up to $\sim 12$ hours before the SN event~\cite{KamLAND:2024uia}.
% An established neutrino alert system, the $\mathrm{SN}$ Early Warning System (SNEWS)~\cite{Antonioli:2004zb}, is already in place. 
% SNEWS provides an early warning for a Galactic CCSN by tracking coincidences between neutrino experiments. 
% Its successor, SNEWS 2.0~\cite{SNEWS:2020tbu}, will reduce alert latency, providing faster notifications of potential SN events. 
% Furthermore, the upgraded system will integrate new types of detectors, including large dark matter detectors, which can provide additional data on neutrino fluxes.
% 
Extracting directional information is more challenging.
A detailed study of the potential to accurately detect the position of the source from pre-SN observations was presented in Ref.~\cite{Mukhopadhyay:2020ubs}.
Information about the upcoming event, including positional data, is expected to be available hours before the event, a time more than sufficient to turn the IAXO telescope towards the star. 
The directional sensitivity of these events is relatively low, due to the weak forward-backward asymmetry of the inverse beta decay process used to detect the neutrino events. 
For current liquid scintillators, the low value of this asymmetry translates into a moderate angular resolution of around $60^\circ$ when 200 events are detected, which is realistic for a star like Betelgeuse.
The use of a lithium-loaded liquid scintillator (LS-Li) would substantially increase  the forward-backward asymmetry, pushing the angular resolution to about $15^\circ$ with the same number of events.\footnote{Although currently there is no existing or funded large-volume LS-Li detector specifically optimized for pre-SN neutrino detection, ongoing developments indicate their feasibility. 
Experiments like PROSPECT~\cite{PROSPECT:2018dnc} have successfully demonstrated lithium-loaded scintillator technology at multi-ton scales. 
Additionally, LS-Li detectors are actively being explored in reactor monitoring and neutrino oscillation experiments, such as the Segmented Anti-Neutrino Directional Detector (SANDD) project~\cite{sutanto2021sandd}. 
Furthermore, the conceptual design of the THEIA~\cite{Theia:2019non} detector includes lithium loading among its potential future upgrades, although it remains in a preliminary proposal stage.
Finally, upcoming large-scale neutrino detectors, including DUNE Phase II~\cite{DUNE:2024wvj} and potentially THEIA, are expected to significantly enhance angular resolution capabilities within the operational timeline of IAXO.
} 
Further improvements are possible with proposed future neutrino detectors, such as THEIA~\cite{Theia:2019non}.

\begin{table}[t!]
\renewcommand{\arraystretch}{1.5}
\centering

\begin{tabular}{lccc}
\hline
\hline
$N$ & Time before Supernova & Angular Error (68.8\% C.L.) & Candidates within Cone\\ \hline
5 & $4\,\text{h}$ & $16.44^\circ$ & 2\\ 
5 & $1\,\text{h}$ & $11.16^\circ$ & 1\\ 
5 & $2\,\text{min}$ & $8.54^\circ$ & 1\\ \hline
8 & $4\,\text{h}$ & $23.24^\circ$ & 4\\ 
8 & $1\,\text{h}$ & $15.47^\circ$ & 3\\ 
8 & $2\,\text{min}$ & $11.63^\circ$ & 1-2\\ \hline
13 & $2\,\text{h}$ & $48.26^\circ$ & 7\\ 
13 & $1\,\text{h}$ & $19.00^\circ$ & 2-3\\ 
13 & $2\,\text{min}$ & $9.84^\circ$ & 1\\ \hline \hline
\end{tabular}
\caption{
Angular error and number of candidates within the error cone over time for three supernovae candidates assuming the usage of LS-Li detectors. 
All three candidates are visible more then 50\% of the time for BabyIAXO. The candidates are identified by their index $N$ from Tab.~\ref{tab:SN-Candidate}. 
Data from Ref.~\cite{Mukhopadhyay:2020ubs}. 
% \GL{What does it mean $1-2$ or $2-3$ candidates? Does it depend on time?}
}
\label{tab:B-IAXO_pos}
\end{table}

Despite the moderate angular precision expected in the case of pre-SN neutrino events, the information provided by an early warning system may suffice to identify the SN candidate, given that the sample of possible close-by candidates is known (see Tab.~\ref{tab:SN-Candidate}), and that their relative angular distances are in most cases relatively large (cf. Tab.~\ref{tab:angular-distances} in Appendix \ref{app:angsep}).
If the progenitor were correctly identified, (Baby)IAXO would have sufficient pointing accuracy to observe it during the collapse. Indeed, the pointing accuracy required for IAXO to follow the Sun—its primary purpose—is $\leq 30^{\prime\prime}$ and the field of view (FOV) associated with a 60–70~cm diameter magnet bore and detector is more than sufficient to enable accurate pointing toward any of the supernova candidates.\footnote{Note that the most stringent pointing requirements for solar axion searches are given by the X-ray telescopes of IAXO. However, such telescopes are not available for MeV photons. Therefore, assuming the detector covers the entire bore, the FOV in this case is estimated from the diameter of the bore divided from the distance of the detector from the bore opening. For standard IAXO geometries, this corresponds to $\approx 2-4$ degrees.}
A more critical issue would arise if multiple potential targets fell within the angular uncertainty cone.
Given the typically large angular separations between supernova sources relative to the IAXO FOV, a decision would be required in such cases.
Since pointing to the wrong object could result in a significant loss of sensitivity, the most natural choice would be to point toward the nearest target, which is expected to yield the strongest signal.
However, a definitive decision-making protocol—adapted to case-specific circumstances—has not been formally discussed or approved by the IAXO collaboration yet.

\begin{figure}[t!]
    \centering
    \includegraphics[width=0.8\linewidth]{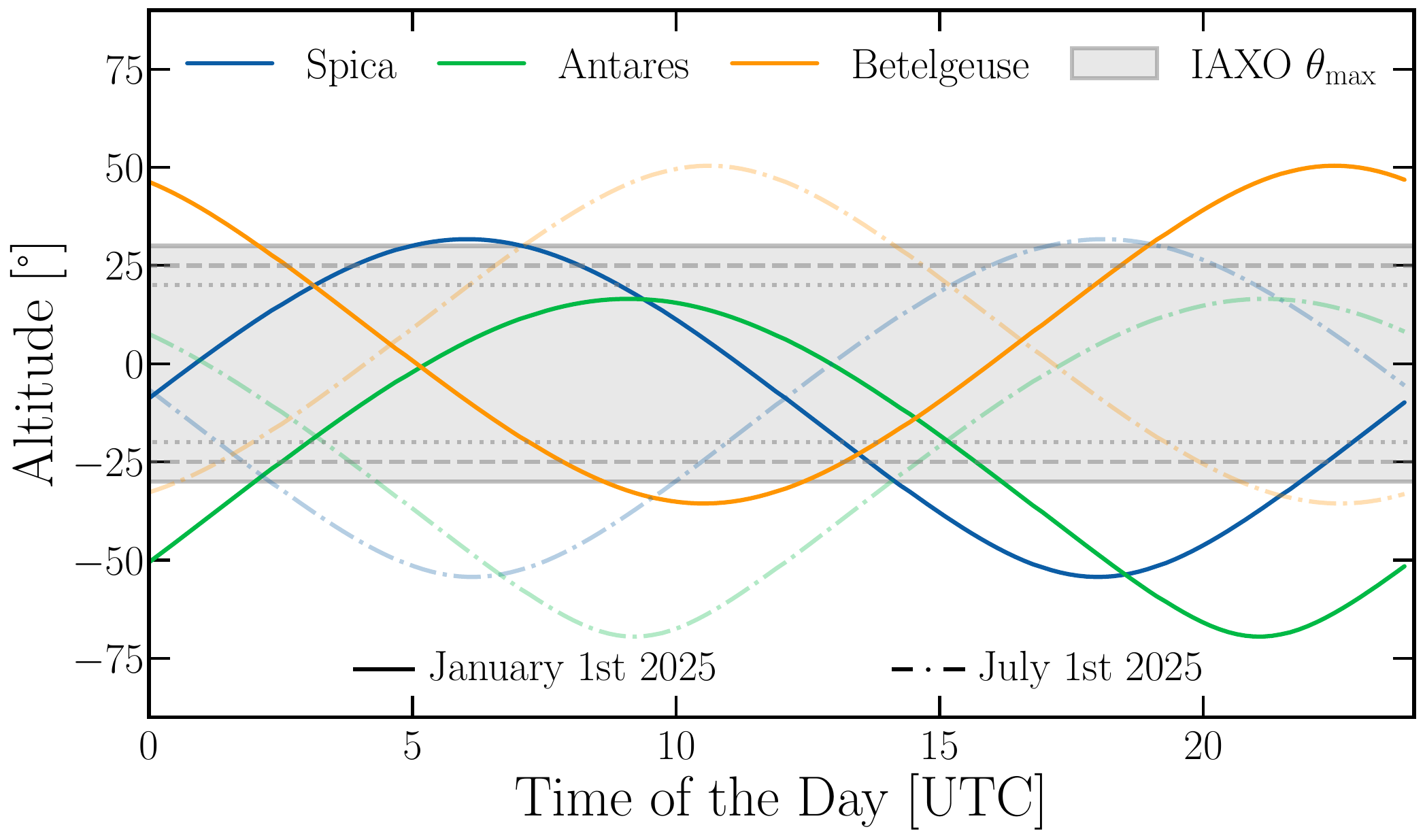}
    \caption{Daily variation of the altitude of three representative SN candidates, Spica, Antares and Betelgeuse, as observed from DESY on the first of January and on the first of July 2025. The gray area highlights the altitudes accessible by IAXO for the three cases $\theta_{\mathrm{max}}$ = 20° (\gray{dotted}), 25° (\gray{dashed}) and 30° (\gray{solid line}) studied in this work, cf. Tab.~\ref{tab:observation-time}.}
    \label{fig:daily_variation}
\end{figure}

An important consideration is the tracking range of IAXO, for which no definite configuration has been determined, aside from the requirement that the altitude angle, measured relative to the horizon, covers a symmetric range that extends at least from $- 21.1^\circ$ to $+ 21.1^\circ$~\cite{IAXO:2019mpb, IAXO:2020wwp}.
For reference, in our analysis we consider three realistic scenarios corresponding to altitude ranges of $\pm 20^\circ$, $\pm25^\circ$, and $\pm 30^\circ$.
In Fig.~\ref{fig:daily_variation}, we show the example of the daily variation of the altitude of three representative SN candidates, Spica, Antares and Betelgeuse, as seen from an observatory  located at DESY (Hamburg, Germany), where (Baby)IAXO will be hosted. 
The figure shows the altitude for two representative days (chosen without any specific reason) to highlight the annual modulation induced by the Earth rotation. 
It is evident that the limited tracking possibilities of the telescope prevent the observation of the star in some period of the day.
We have calculated the percentage of the time during which each star in Tab.~\ref{tab:SN-Candidate} is visible from DESY.\footnote{Notice that the star may not visible in the proper sense, since photons may not reach the telescope, for example if the star is below the horizon. However, here we mean \textit{\enquote{visible in axions}}.}
The results are shown in Tab.~\ref{tab:observation-time}, where the stars are identified via their index $N$ and common name, as defined in Tab.~\ref{tab:SN-Candidate}.

\begin{table}[t!]
\renewcommand{\arraystretch}{1.3}
\centering
\begin{tabular}{c|l|ccc}
\hline\hline
$N$ & Common Name & \multicolumn{3}{c}{Observable Time Fraction [\%]} \\ 
   &                 & $\theta_{\rm max}=20^\circ$ & $\theta_{\rm max}=25^\circ$ & $\theta_{\rm max}=30^\circ$ \\ \hline
1  & Spica/$\alpha$ Virginis   & 43  & 63  & 70  \\ %\hline
2  & $\zeta$ Ophiuchi  & 42  & 59  & 71  \\ %\hline
3  & $\alpha$ Lupi     & 28  & 36  & 43  \\ %\hline
4  & $\lambda$ Velorum & 33  & 40  & 46  \\ %\hline
5  & Antares/$\alpha$ Scorpii  & 49  & 54  & 59  \\ %\hline
6  & $\epsilon$ Pegasi & 42  & 57  & 71  \\ %\hline
7  & $\beta$ Arae      & 9   & 24  & 34  \\ %\hline
8  & Betelgeuse/$\alpha$ Orionis & 41  & 54  & 74  \\ %\hline
9  & q Car/V337 Car    & 0   & 0   & 23  \\ %\hline
10 & $\zeta$ Cephei    & 0   & 18  & 30  \\ %\hline
11 & Rigel/$\beta$ Orionis     & 41  & 55  & 73  \\ %\hline
12 & $\zeta$ Cygni     & 46  & 51  & 56  \\ %\hline
13 & S Monocerotis A   & 42  & 57  & 71  \\ %\hline
14 & S Monocerotis B   & 42  & 57  & 71  \\ %\hline
15 & q Car/V520 Car    & 0   & 11  & 26  \\ \hline\hline
\end{tabular}
\caption{Potential time fractions for observations of the stars listed in Tab.~\ref{tab:SN-Candidate}, identified by their index $N$ and common names.}
\label{tab:observation-time}
\end{table}

\section{Sensitivity prospects}
\label{sec:prospects}
% \red{JuanAn, Maria}

\begin{figure}[t]
    \centering
    \begin{subfigure}[b]{0.49\textwidth}
        \centering
        \caption{$NN$: Betelgeuse}
        \includegraphics[width=\textwidth]{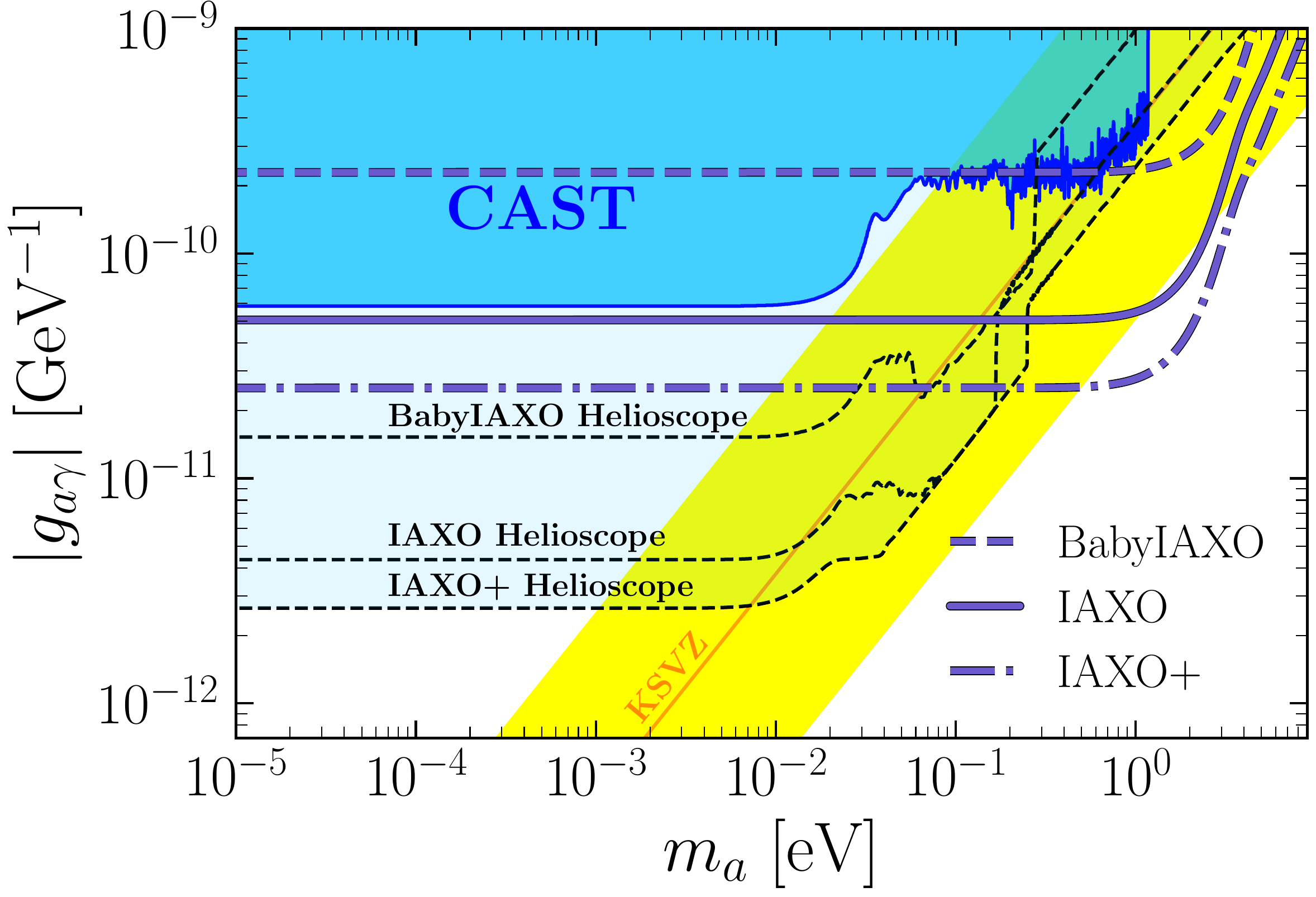}
        \label{}
    \end{subfigure}
    \hfill
    \begin{subfigure}[b]{0.49\textwidth}
        \centering
        \caption{$NN$: Spica}
        \includegraphics[width=\textwidth]{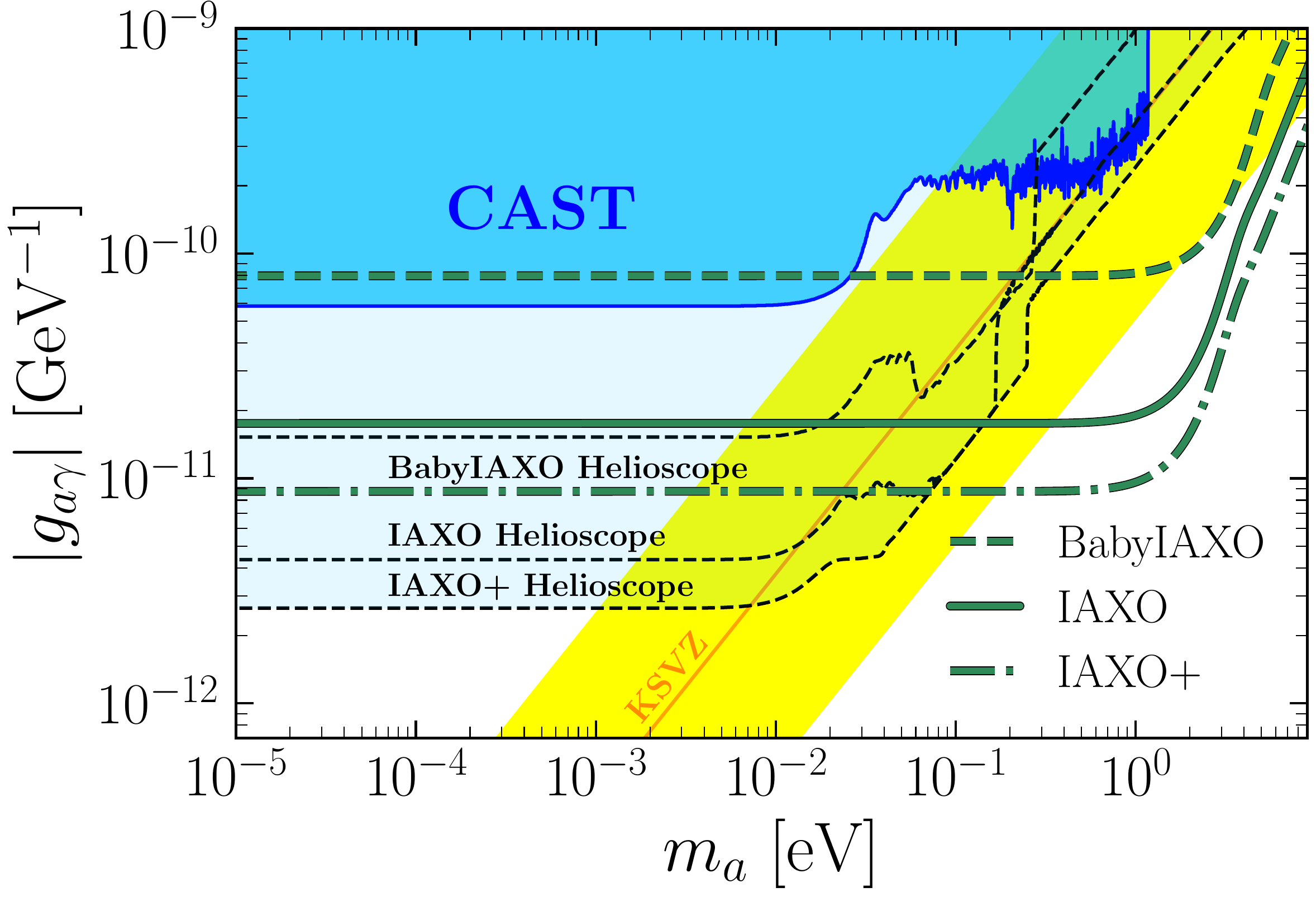}
        \label{}
    \end{subfigure}
    \vfill
    \begin{subfigure}[b]{0.49\textwidth}
        \centering
        \caption{$NN +\pi N$: Betelgeuse}
        \includegraphics[width=\textwidth]{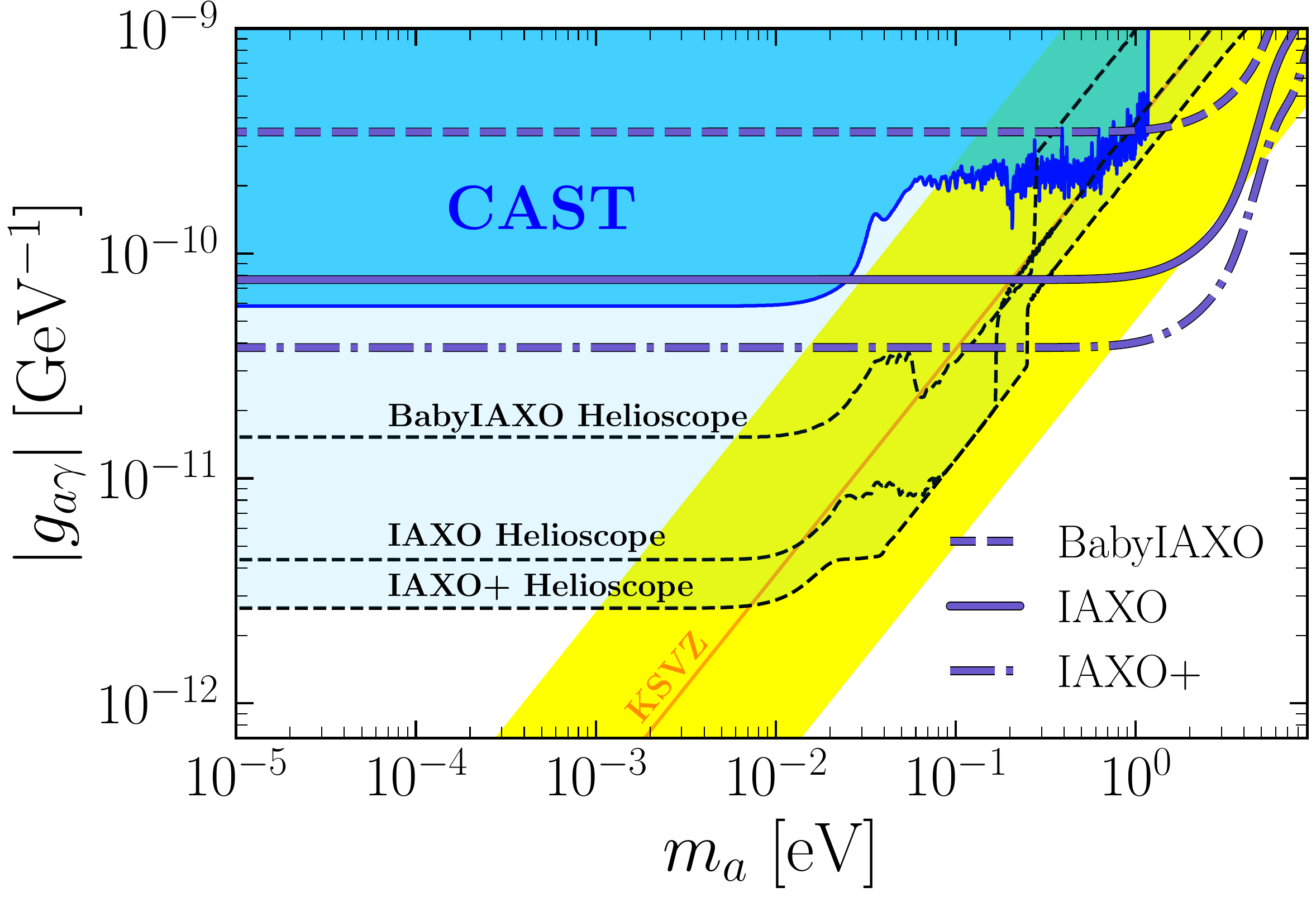}
        \label{}
    \end{subfigure}
    \hfill
    \begin{subfigure}[b]{0.49\textwidth}
        \centering
        \caption{$NN +\pi N$: Spica}
        \includegraphics[width=\textwidth]{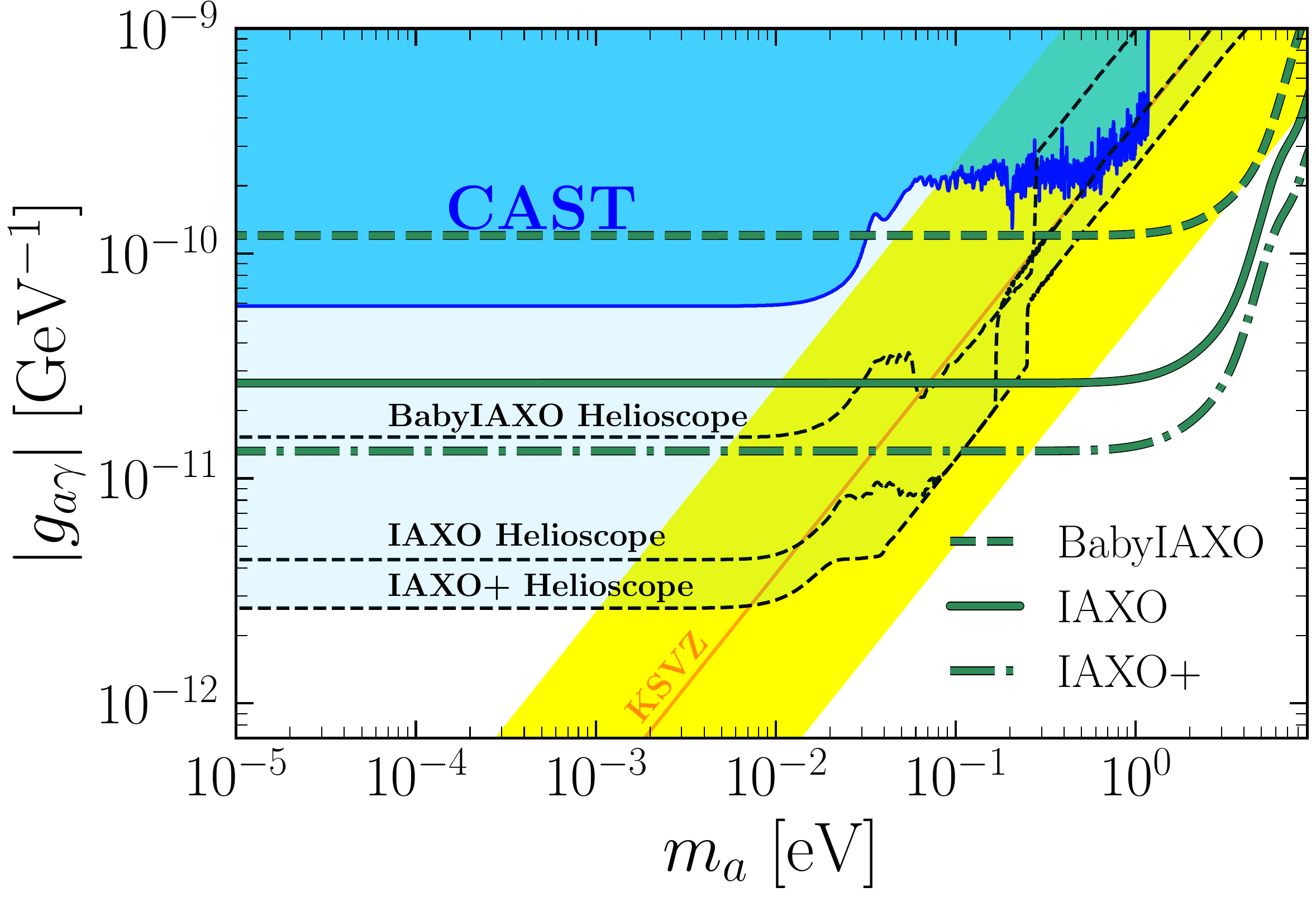}
        \label{}
    \end{subfigure}
    \caption{Sensitivity prospects for BabyIAXO, IAXO and IAXO+ as SNScopes for Betelgeuse (violet, \textit{left panel}) and Spica (green, \textit{right panel}) in the ALP free streaming regime considering bremsstrahlung emission and both bremsstrahlung and pion conversion emission. In the upper panels, we set the ALP-nucleon coupling to ${g_{aN} = 10^{-9}}$, while in the lower panels we set it to the maximum value allowed by the SN cooling argument ${g_{aN} = 5\times10^{-10}}$. The black transparent regions show the sensivity prospects of IAXO, BabyIAXO and IAXO+ as helioscopes. 
    }
    \label{fig:AxionPhoton_free_streaming}
\end{figure}

\begin{figure}[t]
    \centering
    \begin{subfigure}[b]{0.49\textwidth}
        \centering
        \caption{Trapping: Betelgeuse}
        \includegraphics[width=\textwidth]{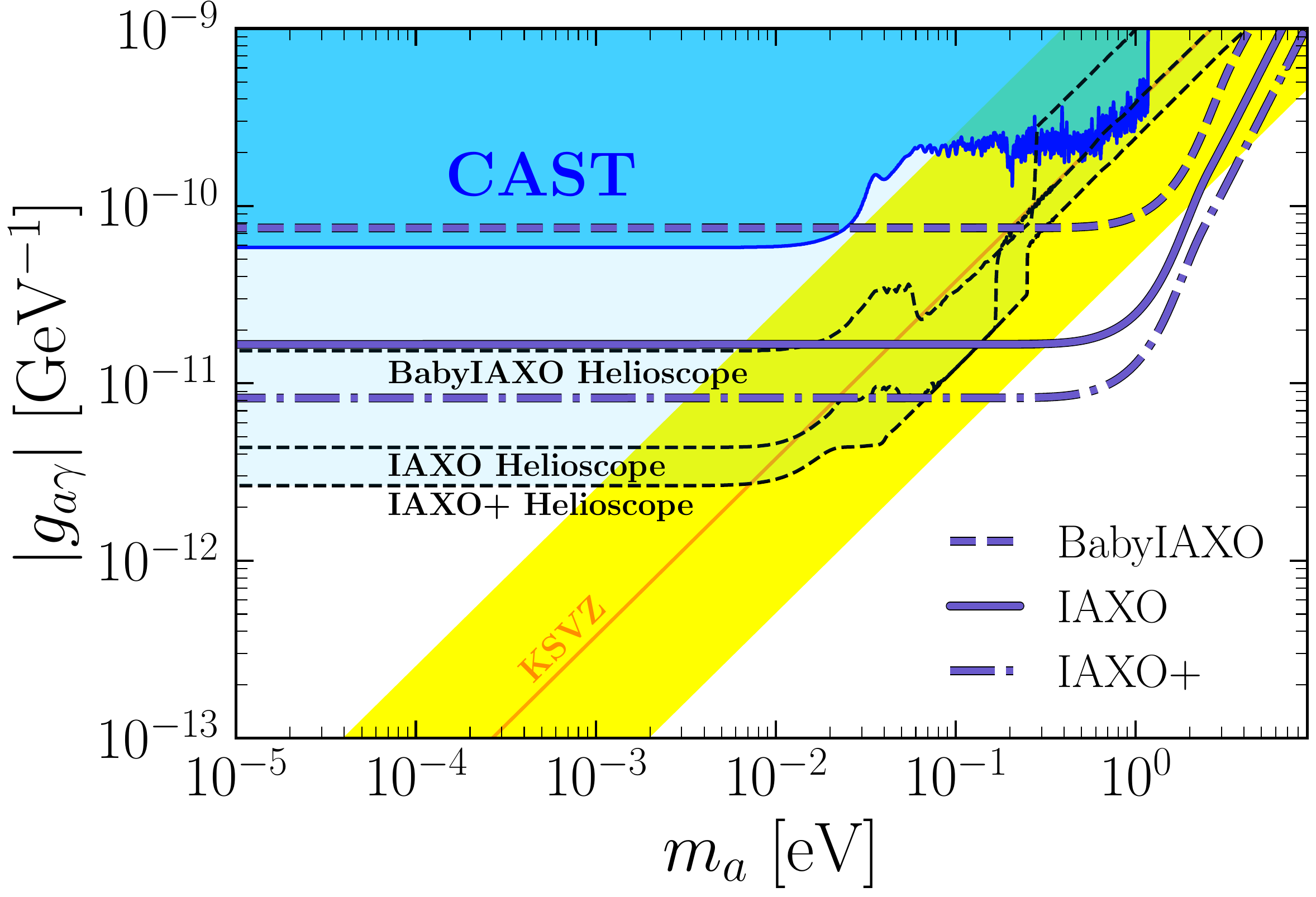}
        \label{}
    \end{subfigure}
    \hfill
    \begin{subfigure}[b]{0.49\textwidth}
        \centering
        \caption{Trapping: Spica}
        \includegraphics[width=\textwidth]{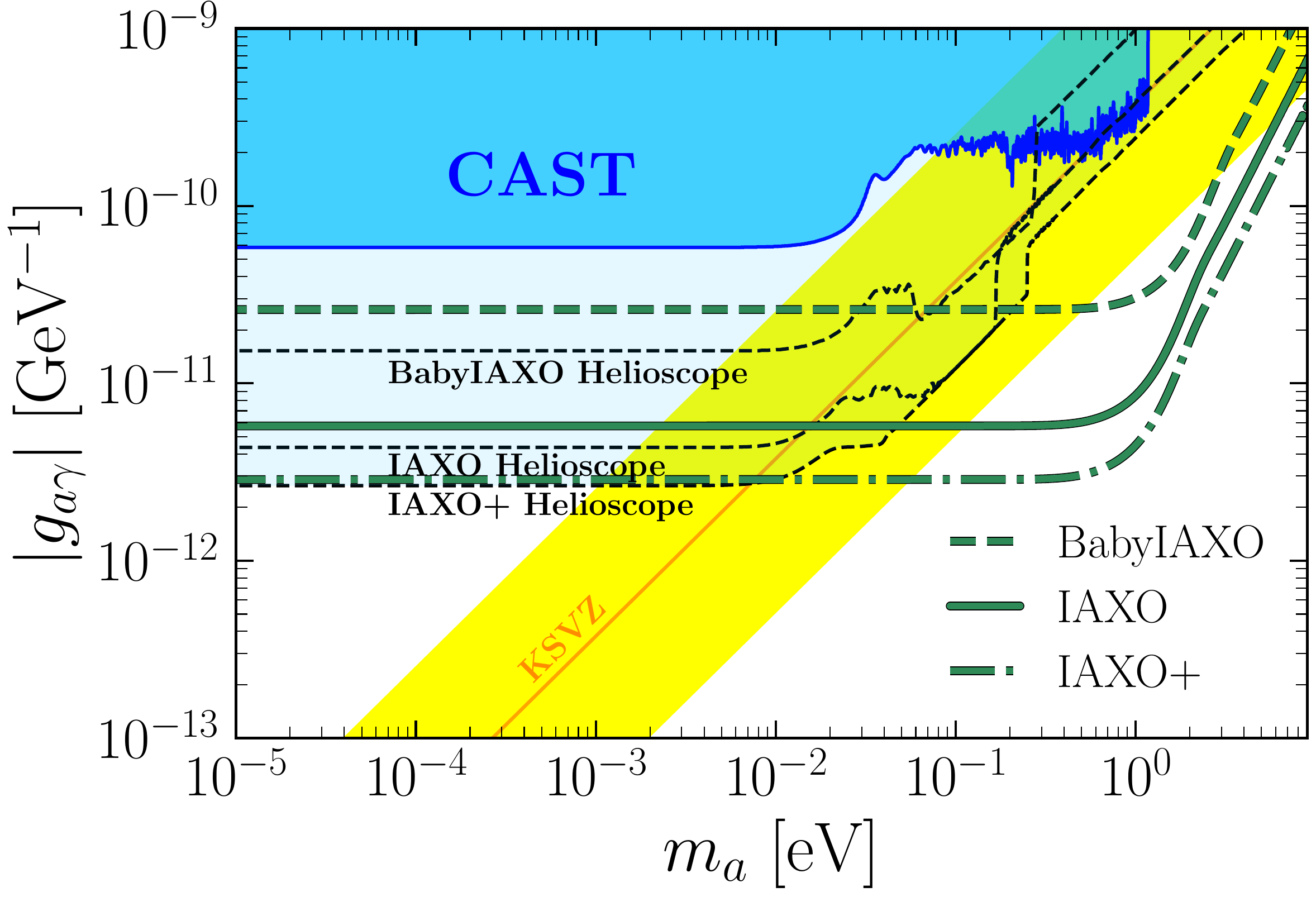}
        \label{}
    \end{subfigure}

    \caption{Sensivity prospects for BabyIAXO, IAXO and IAXO+ as SNScopes for Betelgeuse (violet, \textit{left panel}) and Spica (green, \textit{right panel}) in the trapping regime. Again, here we set the ALP-nucleon coupling to the minimum value allowed by the upper limit of SN cooling bound ${g_{aN} = 3\times10^{-6}}$~\cite{Lella:2023bfb}. The blue transparent regions show the sensivity prospects of IAXO, BabyIAXO and IAXO+ as helioscopes.%\mg{Make sure the CAST bound shown is the latest one (submitted PRL) not the Nature 2017.}
    }
    \label{fig:AxionPhoton_trapping}
\end{figure}

Axions could be abundantly produced during a core-collapse SN event with energies of about 10–250 MeV, as shown in Fig.~\ref{fig:AxSpectra}. 
The probability of axion to photon conversion inside an approximately uniform magnetic field $B$ with a length $L$ can be reduced to~\cite{Raffelt:1987im,VanBibber:1989}

\begin{equation}\label{eq:ConversionProb}
P_{a\rightarrow\gamma} = \left (\frac {g_{a\gamma} B L}{2} \right)^2 \left [\frac{\sin \left(\frac{qL}{2} \right)}{\frac{qL}{2} } \right ]^2,
\end{equation}
where $ g_{a\gamma} $ is the axion-to-photon coupling constant and $q$ is the momentum transfer between the axion and the photon.
In the relativistic limit, in vacuum, 
\begin{equation}\label{eq:momentumTrans}
q = \frac{m_{a}^2 }{2E_{a}},
\end{equation}
where $m_{a}$ is the axion mass and $E_{a}$ the axion energy.
Therefore, the number of expected signal counts $S_{\gamma}$ {associated to the ALP emission during the SN cooling phase lasting $\sim10\,\s$}, is given by
% \begin{equation}\label{eq:signal}
% S_{\gamma} = \int_{E_{i}}^{E_{f}} \mathrm{d}E\,\frac {\mathrm{d} \Phi_{a}^{\mathrm{SN}} } {\mathrm{d} E}\,P_{a\rightarrow\gamma} \,\epsilon\,A = g_{a\gamma}^2\,g_{aN}^2 \int_{E_{i}}^{E_{f}} \mathrm{d}E\,\frac {\mathrm{d}n_{\gamma}} {\mathrm{d} E} ,
% \end{equation}
\begin{equation}\label{eq:signal}
S_{\gamma} =\frac{1}{4\pi D^2} \int_{E_{i}}^{E_{f}} \mathrm{d}E\,
\frac {\mathrm{d} N_{a} } {\mathrm{d} E}\,P_{a\rightarrow\gamma} \,\epsilon\,A = g_{a\gamma}^2\,g_{aN}^2 \int_{E_{i}}^{E_{f}} \mathrm{d}E\,\frac {\mathrm{d}n_{\gamma}} {\mathrm{d} E} ,
\end{equation}
where ${\mathrm{d} N_{a} }/ {\mathrm{d}E}$ is the emitted axion spectrum integrated over $\sim10\,\s$, defined in Eq.~\eqref{eq:emittd_axion_spectrum-FreeStreaming}; 
$D$ is the SN distance;
$P_{a\rightarrow\gamma}$ is the conversion probability of the axion into a photon in a uniform magnetic field, presented in Eq.~\eqref{eq:ConversionProb}; 
$\epsilon$ is the detector efficiency and $A$ is the magnet bore area. 
The second identity of Eq.~\eqref{eq:signal}, together with Eq.~\eqref{eq:emittd_axion_spectrum-FreeStreaming}, define the quantity $dn_\gamma/dE$, which parameterizes (though it is not exactly equal to) the number counts per energy.
% \sout{In the second identity of Eq.~\eqref{eq:signal} the formula has been reduced. 
% As shown, the expected number of counts is proportional to $g_{a\gamma}^2 \times g_{aN}^2$, while ${\mathrm{d} n_{\gamma}}/{\mathrm{d} E}$ is the number of signal counts per energy unit, which is given by:}
% \mg{Erase the equation below.}
% \begin{equation}\label{eq:signalCounts}
% \frac {\mathrm{d} n_{\gamma}} {\mathrm{d} E} = \frac {\mathrm{d} N_{a}^{\mathrm{SN}}} {\mathrm{d} E} \frac {1} {4\pi D^2} \epsilon\,A\,\left (\frac {B L}{2} \right)^2 \left [\frac{\sin \left(\frac{qL}{2} \right)}{\frac{qL}{2} } \right ]^2,
% \end{equation}
% \sout{where ${\mathrm{d}N_{a}^{\mathrm{SN}} }/{\mathrm{d}E}$ is the differential SN axion emission spectrum integrated over the $10\,\s$ introduced in Sec.~\ref{sec:axion_prod} and $D$ is the SN distance.} 
For future convenience, it is useful to split this function into the bremsstrahlung and pion parts:
\begin{equation}
\label{eq:n_gamma_splitted}
    n_{\gamma} = n_{\gamma}^{NN} + \delta\cdot n_{\gamma}^{\pi N}\,,
\end{equation}
which follows our definition in Eq.~\eqref{eq:emittd_axion_spectrum-FreeStreaming}.
Under the assumption of zero detected counts and zero background expected counts, the exclusion limit at a 95\% of C.L. is given by:
\begin{equation}\label{eq:limit}
%g_{a\gamma}^2 \times g_{aN}^2 \leq {\frac{\operatorname{ln}(20)}{ S_{\gamma} } }.
S_{\gamma} = g_{a\gamma}^2\,g_{aN}^2 \int_{E_{i}}^{E_{f}} \mathrm{d}E\,\frac {\mathrm{d}n_{\gamma}} {\mathrm{d}E} \leq \operatorname{ln}(20)\,.
\end{equation}
% 
%\maria{Based on GEANT4 Montecarlo simulations, in this work we are considering an efficiency of 95\% for high energy gammas for the LiquidO-based detector (María: or maybe only say that we are considering a 95\% efficiency, without mentioning GEANT4). Also, we set a conservative energy threshold of 3 MeV (E$_i$).}
Sensitivity prospects are shown in Fig. \ref{fig:AxionPhoton_free_streaming} and \ref{fig:AxionPhoton_trapping} for two different SN progenitor candidates, namely Spica ($\alpha$ Virginis) and Betelgeuse ($\alpha$ Orionis).\footnote{We underline that the value of the ALP-nucleon coupling assumed in the trapping regime, $g_{aN} = 3 \times 10^{-6}$, is disfavored, as it would have led to axion-induced events in Kamiokande II coinciding with the SN 1987A neutrino burst (see Refs.~\cite{Engel:1990zd,Lella:2023bfb}). Nevertheless, searches for supernova axions with IAXO would still offer a valuable and complementary probe of this region of parameter space.}
These sensitivities have been computed in different helioscope scenarios:
BabyIAXO, consisting of two magnetic bores, with a length $L=10$~m and an area $A=0.77$~m$^2$ as well as an average magnetic field $B=2$~T; IAXO, which will be equipped with 8 magnet bores with $L=20$~m and $A=2.3$~m$^2$, providing a magnetic field of strength $B=2.5$~T; and IAXO+, which represents an enhanced version of IAXO, with a magnetic field $B=3.5$~T, an area $A=3.9$~m$^2$ and magnet bores with $L=22$~m. 
These helioscope scenarios are described in Ref.~\cite{IAXO:2020wwp} and summarized in Table \ref{tab:iaxo_parameters}. For the high-energy gamma detector we assume a conservative detection efficiency of $\epsilon = 95\%$ and an energy threshold of 3~MeV.

%\begin{table}[h]
%\renewcommand{\arraystretch}{1.5}
%\centering
%\begin{tabular}{lccc}
%\hline\hline
%       & BabyIAXO & IAXO & IAXO+ \\ \hline
%$B$ (T)  & 2        & 2.5  & 3.5   \\
%$L$ (m)  & 10       & 20   & 22    \\
%$A$ (m$^2$) & 0.77     & 2.3  & 3.9   \\ \hline\hline
%\end{tabular}
%\caption{Magnetic field strength $B$, length $L$, and cross sectional area $A$ for BabyIAXO, IAXO and IAXO+~\cite{IAXO:2020wwp}. \AL{\bf Maybe to make this table more readable it is better to have the dectors on the different rows and their properties on the differen columns. What do you think?}}\label{tab:iaxo_parameters}
%\end{table}
%\maria{Maria: maybe we want this table in section 3?}

\begin{table}[t!]
\renewcommand{\arraystretch}{1.5}
\centering
\begin{tabular}{lccc}
\hline\hline
         & $B$ (T) & $L$ (m) & $A$ (m$^2$) \\ \hline
BabyIAXO & 2     & 10    & 0.77   \\
IAXO     & 2.5   & 20    & 2.3    \\
IAXO+    & 3.5   & 22    & 3.9    \\ \hline\hline
\end{tabular}
\caption{Magnetic field strength $B$, length $L$, and cross sectional area $A$ for BabyIAXO, IAXO and IAXO+~\cite{IAXO:2020wwp}.}\label{tab:iaxo_parameters}
\end{table}

%\maria{[Maria: comment each figure]}
In contrast to helioscope searches, where the sensitivity drops rapidly after 10 meV, the coherence region for SN axions is extended to above 1 eV. 
% {\bf WARNING: FOR 1 eV finite mass effect would imply a time delay on the axion signal.}
% \mg{The time delay is probably very small. 
% It is sub-second in the case of SN 1987A, which is at 50 kpc away. }
This allows for a wider range of the parameter space to be explored, including the theoretically most favoured axion models.

%\maria{[Discussion on the detection case]}
In case of a positive SN axion signature, the energy spectrum of the detected counts will provide precious information about the axion production mechanisms during the core-collapse SN. Here we discuss the signal analysis for the free-streaming regime under the assumption of bremsstrahlung and pion conversion emission. 
At given value of the ALP-nucleon coupling $g_{aN}$, the number of axions emitted via pion conversion processes is highly dependent on the pion abundance in the supernova core, a factor which remains under investigation.
In order to perform a model independent signal analysis, we employ a two-dimensional likelihood analysis in which the probability density function is given by the Poisson distribution. Thus, the likelihood function can be written as
\begin{equation}\label{eq:Like}
L = \frac{1}{L_0}\prod_{i=1}^n e^{-\s_i} \frac{\s_i^{ni}}{n_i!},\qquad L_0 = \prod_{i=1}^n e^{-n_i} \frac{n_i^{ni}}{n_i!},
\end{equation}
where $L_0$ is the normalization factor, the index~$i$ refers to the energy bin, $n_i$ is the number of counts measured in the $i$th bin and $\s_i$ is the expected number of signal counts in the $i$th bin, 
\begin{equation}\label{eq:signalBinned}
\s_i  = g_{a\gamma}^2\,g_{aN}^2 \int_{E_i}^{E_i+\Delta E} \mathrm{d}E\,\frac {\mathrm{d} n_{\gamma}^{NN}} {\mathrm{d} E}\, +\,g_{a\gamma}^2\, g_{aN}^2\,\delta\,\int_{E_i} ^{E_i+\Delta E} \frac {\mathrm{d} n_{\gamma}^{\pi N}} {\mathrm{d} E} \mathrm{d}E,
\end{equation}
with the integral over the signal counts performed between $E_i$ and $E_i+\Delta E$, referring to the corresponding energy bin~$i$ and assuming a bin size of $\Delta E$~=1~MeV.

Instead of using the natural likelihood, it is more convenient to work with the logarithm of the likelihood $\ln{L}$. According to Wilks' theorem~\cite{Wilks:1938}, if certain general conditions are satisfied, the minimum of $-2\ln{L}$ asymptotically approaches a $\chi^2$ distribution
\begin{equation}\label{eq:LogLike}
- \frac{\chi^2}{2} = \ln{L} = \sum_{i=1}^n n_i -\s_i + n_i \log{\frac{\s_i}{n_i}}.
\end{equation}

\begin{figure}[t!]
    \centering
    \includegraphics[width=0.45\textwidth]{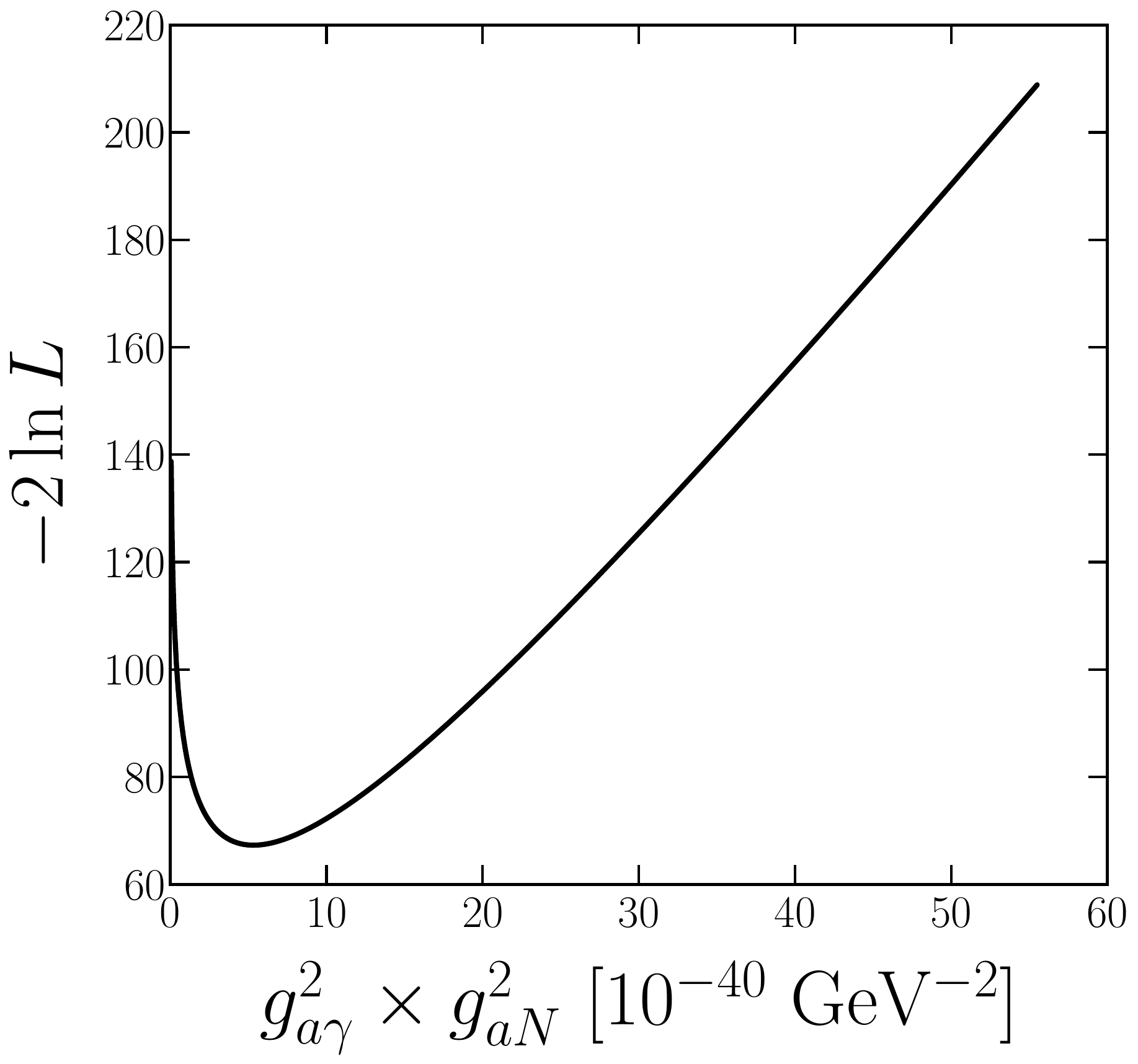}     
    \hfill
    \includegraphics[width=0.45\textwidth]{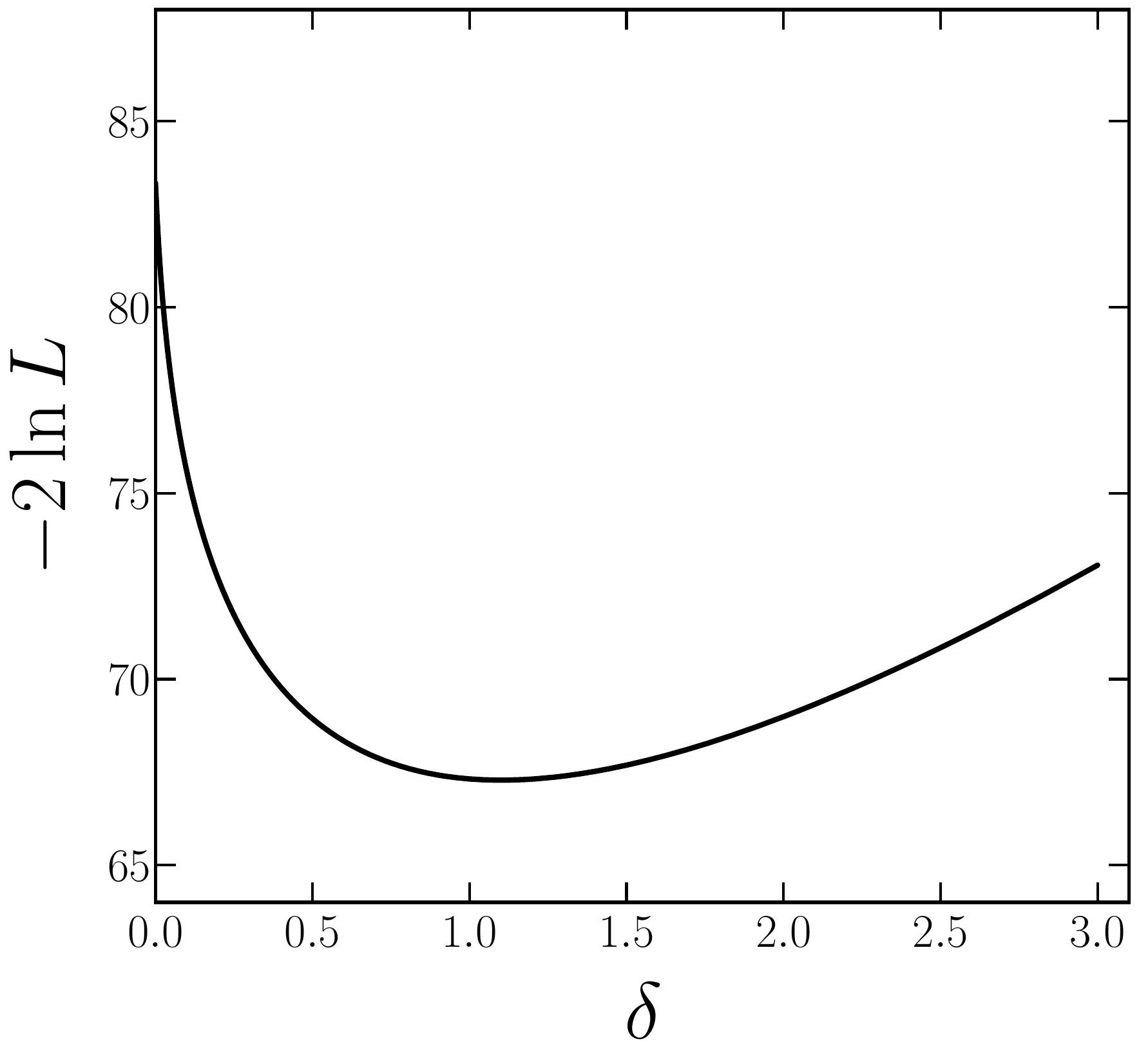} 
   \caption{The $-2\ln{L}$ distribution as a function the different likelihood analysis parameters: $g_{a\gamma}^2 g_{aN}^2$ (left) and the fraction of pions $\delta$ (right).
   }
    \label{fig:LikeAna}
\end{figure}

\begin{figure}[t!]
    \centering
    \includegraphics[width=0.9\linewidth]{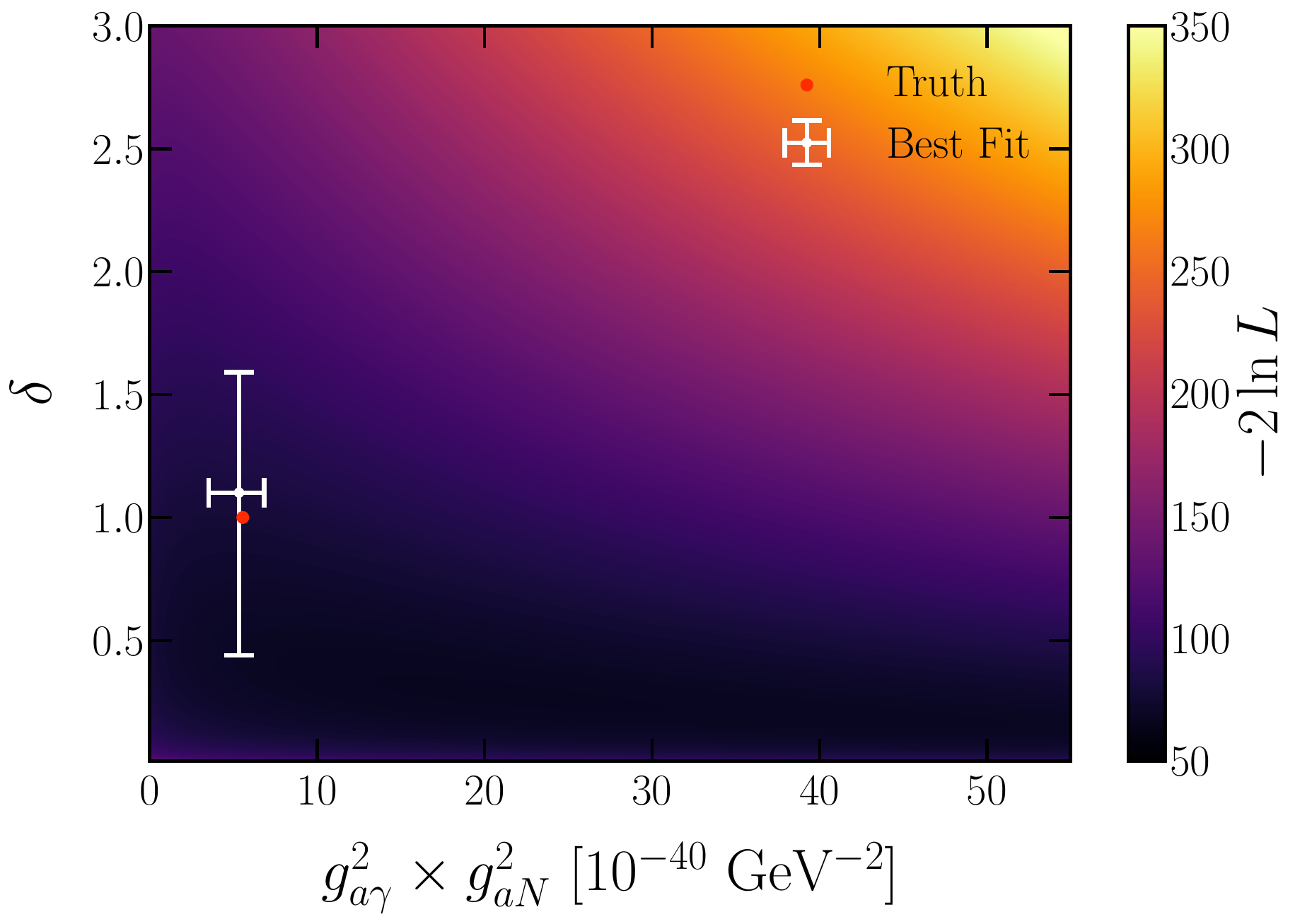}       
   \caption{Likelihood analysis results in the $g_{a\gamma}^2 g_{aN}^2$ and $\delta$  parameter space, in which the color palette shows the value of the $-2\ln{L}$ distribution. The red dot represents the truth value of the SN axion model used as input for the calculations, while the white marker shows the minimum of the likelihood analysis in which the error bars correspond to one standard deviation ($\sigma$).}
    \label{fig:LikeAna2D}
\end{figure}

The likelihood analysis is performed for a fixed axion mass in the coherence region ($m_a = 0.01$~eV), using Eq.~\eqref{eq:LogLike}. Here, $n_i$ represents the number of data counts for a given energy bin and $s_i$ represents the number of the expected counts obtained by integrating Eq.~\eqref{eq:signalBinned} within the same energy bin, scanning over the $g_{a\gamma}^2\times g_{aN}^2$ and $\delta$ parameter space. Mock data are generated using random energy counts distributed according to the free streaming input distribution shown in Fig.~\ref{fig:AxSpectra} with $\delta = 1$ and a coupling constant of $g_{aN} = 5 \times 10^{-10}$. Fig.~\ref{fig:LikeAna} and Fig.~\ref{fig:LikeAna2D} show an illustrative  example of the likelihood analysis results for 10 detected counts. The minimum of the $-2\ln{L}$ distribution is extracted, in which the shift of one $-2\ln{L}$ unit corresponds to one standard deviation ($\sigma$) in the parameter estimation. As shown, the values of the best fit parameters are close to the true values. These results probe the physics potential of IAXO in reconstructing quantities relevant for SN axion emission, if a detection is achieved. 
% \AL{\bf By looking at the plot and from the last sentence, it is not super clear if  $g_{aN}$ is fixed or if it is a free parameter.}
% \AL{\bf Thanks for adjusting the section, now the analysis looks way clearer to me. Maybe a would stress a bit more the fact that we are able to reconstruct the $\delta$ parameter and $g_{a\gamma}^2\times g_{aN}^2$ with good precision. I would write explicitly that this would help us to discriminate if the pion fraction in the core is larger or smaller ($\delta>1$ or $\delta<1$) than what we expect.}
% \mg{I would not specify that we can get $\delta$ with high precision. In the likely case of only a few events we would not be able to measure it very well.}

%%%%%%%%%%%%%%%%%%%%%%%%%%%%%%%%%%%%%%%%%%%%
%%%%%%%%%%%%%%%%%%%%%%%%%%%%%%%%%%%%%%%%%%%%

%%%%%%%%%%%%%%%%%%%%%%%%%%%%%%%%%%%%%%%%%%%%
%%%%%%%%%%%%%%%%%%%%%%%%%%%%%%%%%%%%%%%%%%%%

\section{Discussion and Conclusions}
\label{sec:conclusion}
In this work, we have presented an updated study of the SN axion flux expected from a nearby SN and the corresponding signal it would generate in a helioscope experiment. Our analysis includes axion production from pion-Compton processes, which had not been previously accounted for, despite potentially being the dominant production channel. 
% \sout{a comprehensive analysis of the potential for BabyIAXO and IAXO to detect supernova (SN) axions. 
% Our study underscores the realistic capabilities of these detectors, emphasizing their sensitivity and the scientific opportunities they present.} 
% \mg{THE ABOVE SENTENCE IS NOT TRUE. WE DO NOT DISCUSS THE DETECTORS. WE SHOULD CLARIFY}

We have shown that if a nearby supernova (SN) occurs, IAXO—and, to a somewhat lesser extent, BabyIAXO—could have a realistic chance of detecting QCD axions in regions of parameter space that are inaccessible to other experiments. The parameter space that could be explored extends beyond the stellar bounds derived from globular clusters~\cite{Ayala:2014pea,Straniero:2015nvc,Dolan:2022kul} and even surpasses the regions recently probed through solar observations with NuSTAR~\cite{Ruz:2024gkl}.
Furthermore, our study demonstrates that the area of parameter space accessible to (Baby)IAXO in the case of SN axions extends beyond that explored by standard solar searches of the same instrument. 
This implies that SN axions could, in principle, be discovered by a helioscope before solar axions. 
Consequently, a SN observation offers a genuine discovery opportunity.

Such a detection could provide valuable insights into SN physics, particularly regarding the abundance of negative pions in the SN core.
This would be highly significant, as such a measurement could shed light on the equation of state of matter under extreme conditions.
Experimentally accessing this information is notoriously difficult. While proposals exist in the literature, they primarily focus on very light axion-like particles rather than QCD axions~\cite{Lella:2024hfk}.
In this sense, this work highlights the scientific impact that a future Galactic SN could have for both particle physics and astrophysics, were axions to be detected.

Our study validates the feasibility and scientific value of detecting SN axions with BabyIAXO and IAXO. 
These observations could open new avenues in astrophysics and particle physics, providing insights into the fundamental properties of axions as well as on supernovae, revealing some of their properties which are hard or perhaps impossible to test in other ways. 
A nearby SN event is a very rare possibility and we should be as ready as possible in case such an event were to occur in the next few years. 

\begin{acknowledgments}

%\mg{Igor mentions to add acknowledgments to IAXO ERC \& AEI projects. Can someone add that? }
% \mg{NOTE for the publication Committee: The acknowledgment section needs still a bit of revision, since some of us are moving or have recently changed affiliation. All this will be revised in these days.}

% \mg{Careful: there are two A.L. among the authors. I am using A. Lella for Alessandro Lella. }

This work has been performed as part of the IAXO collaboration.
We warmly thank  Thomas Janka for giving us access to the {\tt GARCHING} group archive.
 This article is based upon work from COST Action COSMIC WISPers (CA21106), 
supported by COST (European Cooperation in Science and Technology). A.Lella kindly thanks the University of Zaragoza for their hospitality and the COST Action COSMIC WISPers (CA21106) for financial support during the visit.\\
The work of PC is supported by the Swedish Research Council under contract 2022-04283.
\\
J.A.G.P, M.G, I.G.I and M.J.P acknowledge support from the European Union’s Horizon 2020 research and innovation programme under the European Research Council (ERC) grant agreement ERC-2017-AdG788781 (IAXO+).\\
The work of J.A.G.P, M.G, I.G.I, and M.J.P is also supported by grant PID2019-108122GB-C31, funded by MCIN/AEI/10.13039/501100011033 and grant PID2022-137268NB-C51, funded by MCIN/AEI/10.13039/501100011033/FEDER, as well as funds from “European Union NextGenerationEU/PRTR” (Planes complementarios, Programa de Astrof\'isica y F\'isica de Altas Energ\'ias).\\
J.A.G.P acknowledges support from the European Union’s Horizon 2020
research and innovation programme under the Marie Skłodowska-Curie grant agreement No 101026819 (LOBRES) as well as the Agencia Estatal de Investigación (AEI) under grant EUR2023-143444, funded by MCIN/AEI/10.13039/501100011033 and by the "European Union NextGenerationEU/PRTR". \\
Additionally, the work of M.G and M.K is supported by the grant PGC2022-126078NB-C21 funded by MCIN/AEI/ 10.13039/501100011033 and “ERDF A way of making Europe”. M. G and M.K further acknowledge the grants DGA-FSE 2023-E21-23R and DGA-FSE 2020-E21-17R, respectively. Both are funded by the Aragon Government and the European Union – Next Generation EU Recovery and Resilience program on `Astrofísica y Física de Altas Energías' CEFCA-CAPA-ITAINNOVA.\\
M.J.P and M.K are further supported by the Government of Aragón, Spain, with PhD fellowships as specified in ORDEN CUS/621/2023 and ORDEN CUS/702/2022, respectively. 
The work of A.M. and A.Lella was partially supported by the research grant number 2022E2J4RK ``PANTHEON: Perspectives in Astroparticle and
Neutrino THEory with Old and New messengers" under the program PRIN 2022 funded by the Italian Ministero dell’Universit\`a e della Ricerca (MUR). 
This work is (partially) supported
by ICSC – Centro Nazionale di Ricerca in High Performance Computing.\\
A.L acknowledges support by the Deutsche Forschungsgemeinschaft (DFG, German Research Foundation) under Germany’s Excellence Strategy – EXC 2121 ``Quantum Universe" – 390833306.\\
G.L acknowledges support from the U.S. Department of Energy under contract number DE-AC02-76SF00515 and, when this work was started, the EU for support via ITN HIDDEN (No 860881). \\

\end{acknowledgments}
%%%%%%%%%%%%%%%%%%%%%%%%%%%%%%%%%%%%%%%%%%%%%%%%%%%%%%%%%%%%

\bibliographystyle{bibi.bst}
\bibliography{references.bib}

\appendix

\section{SN models}
\label{app:SNmodels}
The results discussed in this work are obtained using as a benchmark the state-of-the-art 1D spherical-symmetric {\tt GARCHING} group's SN model SFHo-s18.8 provided in Ref.~\cite{SNarchive}, already used in Refs.~\cite{Bollig:2020xdr,Caputo:2021rux,Caputo:2022mah,Lella:2022uwi,Lella:2023bfb,Lella:2024hfk,Lella:2024dmx}. This model, launched from a stellar progenitor with mass $18.8~M_\odot$~\cite{Sukhbold:2017cnt} and leading to a NS with baryonic mass $1.35~M_\odot$, is based on the neutrino-hydrodynamics code 
{\tt PROMETHEUS-VERTEX}~\cite{Rampp:2002bq}. As further discussed in Ref.~\cite{Fiorillo:2023frv}, the code takes into account all neutrino reactions relevant for core-collapse SNe~\cite{Buras:2005rp,Janka:2012wk,Bollig:2017lki}, and includes a 1D treatment of PNS convection via a mixing-length description of the convective fluxes~\cite{Mirizzi:2015eza} as well as muon physics~\cite{Bollig:2017lki}. On the other hand, current state-of-the-art simulations do not include pions, since their properties in the hot PNS are still under debate. Thus, the presence of pions in the SN core has been taken into account by following the procedure illustrated in Ref.~\cite{Fischer:2021jfm}, including the pion-nucleon interaction as described in Ref.~\cite{Fore:2019wib}.

%%%%%%%%%%%%%%%%%%%%%%%%%%%%%%%%%%%%%%%%%%%%
\begin{figure}[]
    \centering
    \includegraphics[width=0.7\linewidth]{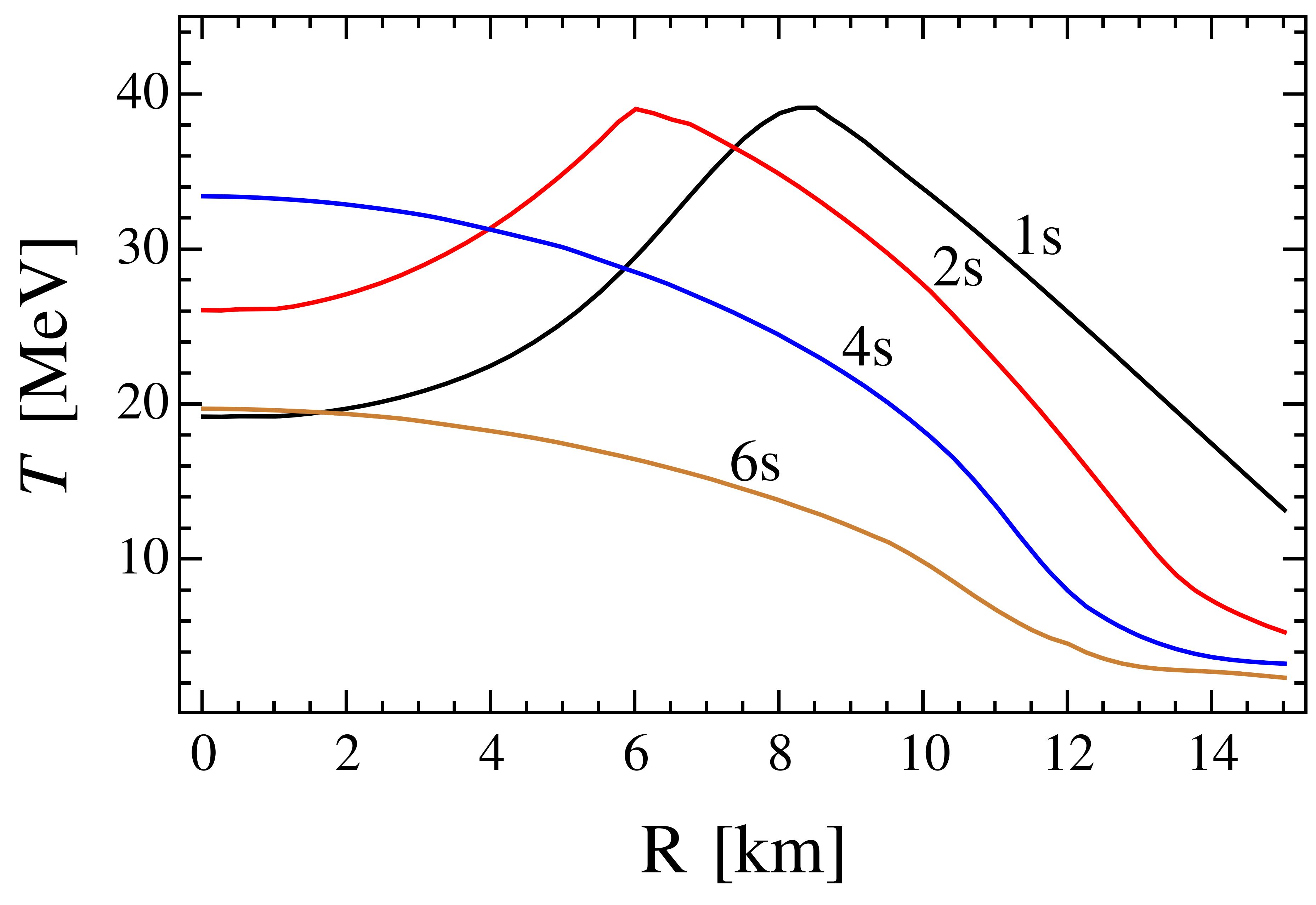}       
    \includegraphics[width=0.7\linewidth]{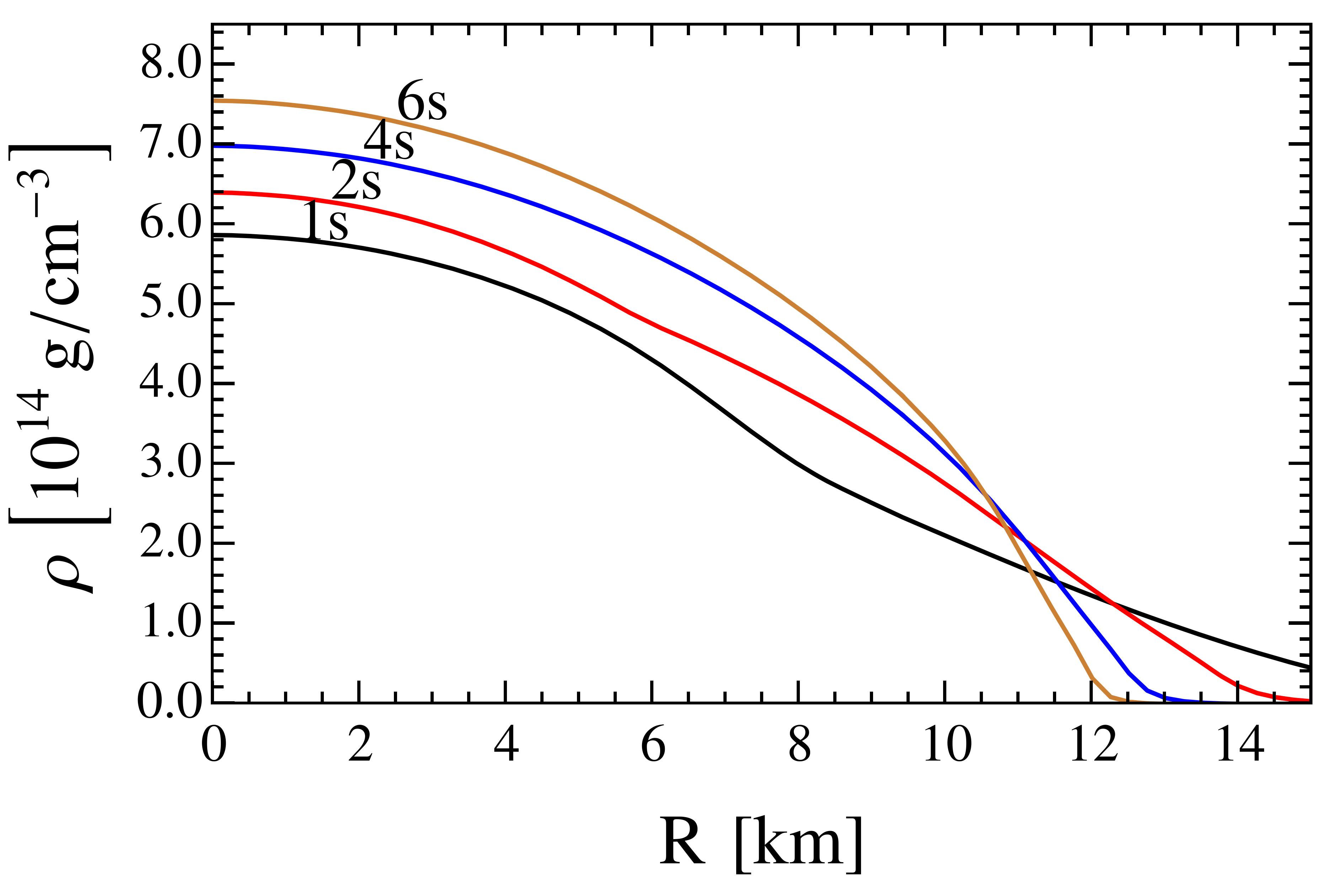}    
    \includegraphics[width=0.7\linewidth]{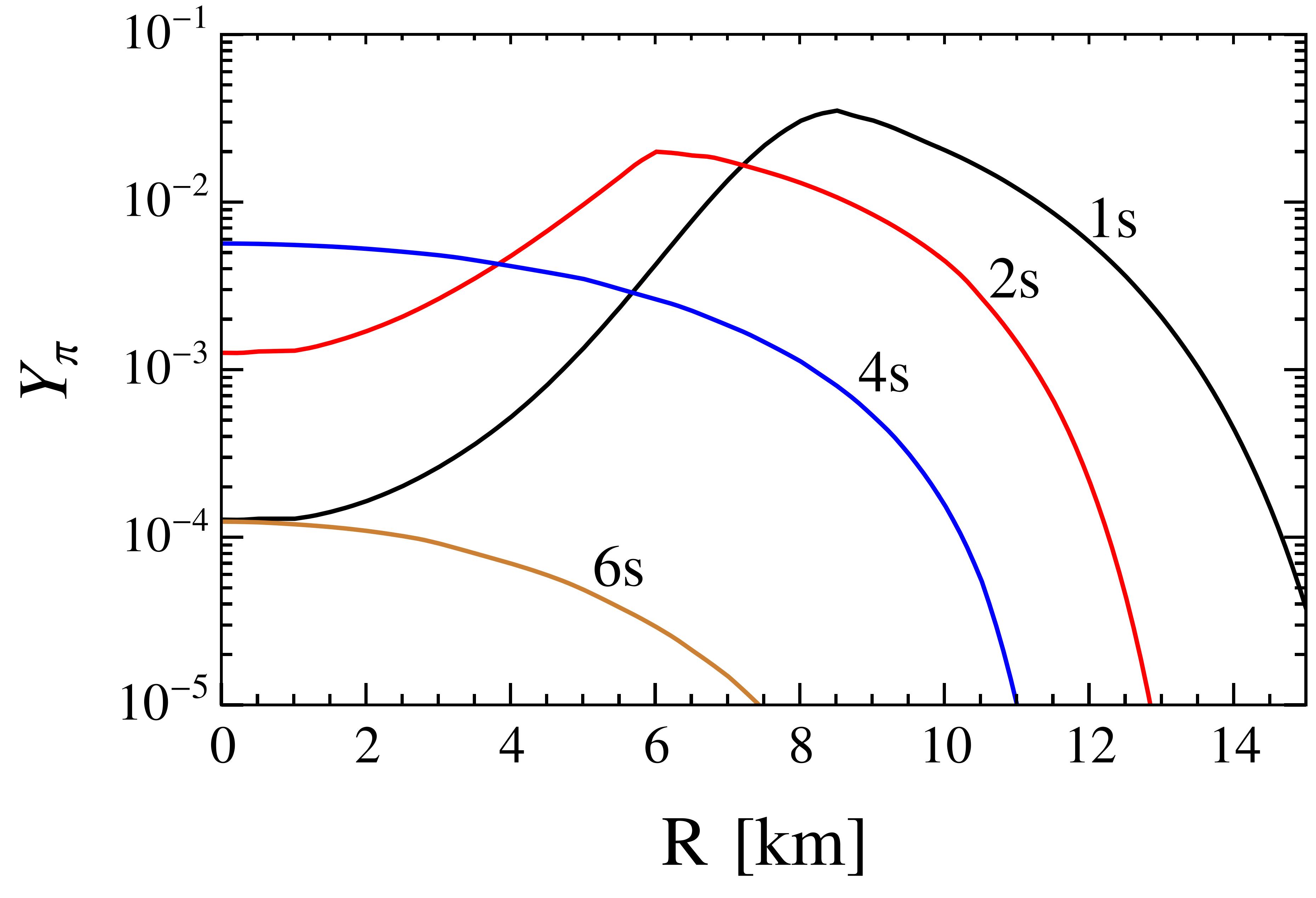}    
   \caption{Radial profiles of the temperature $T$ (upper panel), density $\rho$ (central panel) and pion fraction $Y_\pi$ (lower panel) for our benchmark SN model SFHo-s18.8 at different post-bounce times.}
    \label{fig:Trho}
\end{figure}
%%%%%%%%%%%%%%%%%%%%%%%%%%%%%%%%%%%%%%%%%%%%

We show in Fig.~\ref{fig:Trho} the radial profile of the temperature (upper panel), density (central panel) and pion fraction (lower panel), at post-bounce times $t=1$~s (black), $t=2$~s (red), $t=4$~s (blue) and $t=6$~s (brown) for our benchmark model. In the early-cooling phase, at $t=1-2$~s, the highest temperature ($T_{\rm max}\approx 40$~MeV) is reached at $R\approx 6-9$~km, while at late time ($t \gtrsim 4$~s) the temperature is peaked in the center ($T_{\rm max}\approx 32$~MeV at $t=4$~s and $T_{\rm max}\approx 20$~MeV at $t=6$~s) and decreases at larger radii. On the other hand, due to the protoneutron star contraction, the inner-core density monotonically increases from $\rho_{\rm max} \approx 6\times 10^{14}~{\rm g}/\cm^3$ at $t=1$~s to $\rho_{\rm max} \approx 8\times 10^{14}~{\rm g}/\cm^3$ at $t=6$~s. Finally, the pion fraction peaks in coincidence to the temperature peak and follows the time behavior of the temperature profile. In particular, the highest pion abundances in the core are expected to be
at $t=1-2$~s, where they can reach values as high as $Y_\pi\sim2-3\,\%$, while they are significantly suppressed at later times. These conditions lead to the time evolution of the axion emission spectrum for our benchmark model, described in Appendix~A.1 of Ref.~\cite{Lella:2024hfk}. In this work, we perform our analysis using the axion emission spectrum integrated over 10~s after the core bounce.

\section{Angular Separations between SN Candidates}
\label{app:angsep}
In this appendix we give details on the angular separation between all the stars presented in Tab.~\ref{tab:SN-Candidate}.
The stars are identified by their index, defined in column $N$.
The angular distance matrix is evidently symmetric
\begin{equation}
    \Delta\Theta_{ij} = \begin{cases}
    0 & \text{if } i = j, \\
    \Delta\theta_{ij} & \text{if } i < j, \\
    \Delta\theta_{ji} & \text{if } i > j,
\end{cases}
\end{equation}
where $\Delta\theta_{ij} = \Delta\theta_{ji}$. For the SN candidates considered in this work (cf. Table \ref{tab:SN-Candidate}), the separation angles are shown in Tab.~\ref{tab:angular-distances}, 
where $i \in \left(1,\dots,14\right)$, $j \in \left(2,\dots,15\right)$. 
The minimal angular separation, i.e. the angular distance to the closest neighbour is highlighted.
\begin{table}[h]
\centering
\caption{Angular distances between all the stars in Tab.~\ref{tab:SN-Candidate}. 
The stars are identified by the $i,~j$ indexes. 
Colored data indicate the smallest angle separation associated with each star.}
\footnotesize{
\begin{tabular}{c|*{15}{c}}
$\Delta\theta_{ij}$ & 2 & 3 & 4 & 5 & 6 & 7 & 8 & 9 & 10 & 11 & 12 & 13 & 14 & 15 \\
\hline
1 &  47.1 & \textcolor{teal}{39.7} & 64.8 & 45.9 & 125.7 & 83.2 & 113.4 & 60.5 & 120.4 & 119.8 & 120.3 & 102.6 & 102.6 & 57.3 \\
2 &  & 44.1 & 99.3 & \textcolor{teal}{16.0} & 79.0 & 66.9 & 160.5 & 82.6 & 95.5 & 159.1 & 79.1 & 149.5 & 149.5 & 80.1 \\
3 &  &  & 56.6 & \textcolor{teal}{29.7} & 107.8 & 108.5 & 122.8 & 38.5 & 139.5 & 115.0 & 117.8 & 117.5 & 117.5 & 36.0 \\
4 &  &  &  & 86.3 & 145.1 & 143.9 & 66.2 & \textcolor{teal}{21.7} & 161.9 & 60.6 & 166.3 & 62.3 & 62.3 & 22.6 \\
5 &  &  &  &  & 84.5 & 82.8 & 152.1 & 67.8 & 109.9 & 143.7 & 89.6 & 144.8 & 144.8 & \textcolor{teal}{65.6} \\
6 &  &  &  &  &  & 67.7 & 120.4 & 127.6 & 48.6 & 113.5 & \textcolor{teal}{21.1} & 130.3 & 130.3 & 128.6 \\
7 &  &  &  &  &  &  & 116.8 & 143.2 & \textcolor{teal}{37.3} & 132.6 & 48.0 & 112.7 & 112.7 & 140.0 \\
8 &  &  &  &  &  &  &  & 85.8 & 96.9 & 18.6 & 118.2 & \textcolor{teal}{11.6} & \textcolor{teal}{11.6} & 87.7 \\
9 &  &  &  &  &  &  &  &  & 176.2 & 76.5 & 146.9 & 83.4 & 83.4 & \textcolor{teal}{3.2} \\
10 &  &  &  &  &  &  &  &  &  & 105.3 & \textcolor{teal}{29.3} & 99.8 & 99.8 & 175.4 \\
11 &  &  &  &  &  &  &  &  &  &  & 118.8 & \textcolor{teal}{28.1} & \textcolor{teal}{28.1} & 79.0 \\
12 &  &  &  &  &  &  &  &  &  &  &  & \textcolor{teal}{124.6} & \textcolor{teal}{124.6} & 147.0 \\
13 &  &  &  &  &  &  &  &  &  &  &  &  & \textcolor{teal}{0.0} & 84.7 \\
14 &  &  &  &  &  &  &  &  &  &  &  &  &  & \textcolor{teal}{84.7} \\
\end{tabular}}
\label{tab:angular-distances}
\end{table}
\clearpage

\end{document}